\documentclass[%
reprint,
superscriptaddress,
preprintnumbers,
 amsmath,amssymb,
 aps,
 floatfix,
 prb,
]{revtex4-2}

\usepackage{graphicx}
\usepackage{amssymb}  
\usepackage{enumitem}

\usepackage{dcolumn}
\usepackage{bm}
\usepackage{lineno}

\usepackage[utf8]{inputenc}
\usepackage[T1]{fontenc}
\usepackage{mathptmx}
\usepackage{etoolbox}
\usepackage{amsmath}

\usepackage{soul,xcolor}

\usepackage{xspace}
\newcommand{\alumina}{Al$_2$O$_3$\xspace}
\usepackage{tikz}
\usetikzlibrary{arrows.meta, positioning}

\begin{document}

\title{From Bulk to Surface: Structure and Dynamics of Amorphous Alumina from Deep Potential Molecular Dynamics}

\author{Zheng Yu}
\thanks{These authors contribute equally to this work}
\affiliation{Department of Chemistry, Princeton University, Princeton, NJ, 08540, United States}
\author{Jiayan Xu}
\thanks{These authors contribute equally to this work}
\affiliation{Department of Chemistry, Princeton University, Princeton, NJ, 08540, United States}
\author{Abhirup Patra}
\email{Abhirup.Patra@shell.com}
\affiliation{Shell International Exploration \& Production Inc., 200 N Dairy Ashford Rd, Houston, Texas 77079, United States}
\author{Sharan Shetty}
\affiliation{Shell India Markets Pvt., Ltd., Mahadeva Kodigehalli, Bengaluru 562149, Karnataka, India}
\author{Detlef Hohl}
\affiliation{Shell Information Technology International Inc., 3333 Highway 6 South, Houston, Texas 77082, United States}
\author{Roberto Car}
\email{rcar@princeton.edu}
\affiliation{Department of Chemistry, Princeton University, Princeton, NJ, 08540, United States}

\begin{abstract}

    Understanding the atomic-scale structure and dynamics of amorphous oxide surfaces is essential for interpreting their chemical reactivity, mechanical stability, and interfacial behavior, yet direct experimental characterization remains challenging. We employ Deep Potential (DP) molecular dynamics to generate large-scale, \textit{ab initio}-quality models of amorphous \alumina bulk glasses and melt-quenched free surfaces, enabling a quantitative analysis of both structure and relaxation dynamics with statistical confidence inaccessible to direct \textit{ab initio} simulation. The trained DP model reproduces experimental liquid and glass structure, captures the cooling-rate dependence of the bulk glass transition, and corrects systematic biases in the polyhedral populations predicted by widely used classical force fields. 
    At the free surface, mass density recovers to bulk values over $\sim$10~\AA{}, while local coordination requires a slightly wider subsurface region to fully converge. The outermost layer is oxygen-enriched, exhibits altered polyhedral connectivity with contracted Al--O bonds, and hosts a broad population of under-coordinated motifs (notably AlO$_3$ and OAl$_2$) whose abundances are governed by glass stability. These reactive Lewis acid and Br\o nsted base sites are locally paired in a manner consistent with bond-valence compensation, yet remain spatially dispersed rather than aggregating into extended clusters. 
    Despite this pronounced structural heterogeneity, the surface relaxes on the same timescale as the bulk and exhibits a comparable glass transition temperature, suggesting that the disordered surface is kinetically stable once formed. Together, these results establish a molecular-level picture of amorphous alumina surfaces and demonstrate the capability of machine-learned potentials to resolve structure--property relationships in disordered oxide interfaces.

\end{abstract}

\maketitle

\section{Introduction}

Amorphous aluminum oxide (a-\alumina) is among the most widely deployed dielectric and protective thin-film materials, with applications spanning microelectronics, catalysis, and energy storage.\cite{wilkHighkGateDielectrics2001,katiyar2005electrical,groner_low-temperature_2004,george_atomic_2010,oviroh_new_2019,youngProbingAtomicScaleStructure2020} 
In these technologies, function is ultimately controlled at the surface---by the coordination environments that define dielectric strength, the acid-base sites that drive catalytic reactivity, and the defect chemistry that governs coating durability.\cite{goldsmith_beyond_2017} 
However, the atomic-scale structure of a-\alumina surfaces remains largely unknown. Unlike their crystalline counterparts, these surfaces lack long-range order, host a distribution of coordination motifs that varies with preparation conditions, and cannot be described by any single surface termination model.

Experimentally, the structure of a-\alumina has been probed using a wide array of spectroscopic, scattering, and microscopy techniques.\cite{lamparterStructureAmorphousAl2O31997,skinner_joint_2013,shiStructureAmorphousDeeply2019,hashimotoStructureAluminaGlass2022,Harper2023ChemSci,harperVibrationalThermalProperties2024, tavakoli2013amorphous} 
These studies have established key average structural features, such as the predominance of four-, five-, and six-fold coordinated Al environments, broad bond-length and bond-angle distributions, and the absence of long-range order.\cite{leeStructureAmorphousAluminum2009,leeStructureDisorderAmorphous2010,saroukanianTemperatureDependent4562013,hashimoto_nmr_2017,baggettoAtomicScaleStructure2017} Surface-sensitive techniques, such as X-ray photoelectron spectroscopy (XPS) and infrared spectroscopy of adsorbed probes, further indicate that a-\alumina surfaces are chemically heterogeneous and host under-coordinated sites that strongly influence adsorption and reactivity.\cite{busca_surface_2014,zhaoNatureFiveCoordinatedGAl2O32022,giacomazzi_infrared_2023} However, most measurements provide ensemble-averaged or indirect structural information, making it difficult to resolve spatial variations in coordination, identify subsurface structural gradients, or distinguish distinct local bonding motifs at free surfaces. Moreover, the intrinsic disorder of the amorphous state, which is inherently tied to its thermal history, combined with beam damage, surface contamination, and limited depth resolution, complicates the direct characterization of atomic-scale surface structure.\cite{goldsmith_beyond_2017,rakitaCryoePDFOvercomingElectron2021} As a result, a detailed molecular-level picture of amorphous alumina surfaces, and how this structure depends on thermal history and glass stability, remains incomplete, strongly motivating complementary insights from atomistic simulation.

A central challenge in modeling amorphous materials is satisfying the simultaneous need for computational efficiency and \textit{ab initio}-level accuracy.\cite{friederich2021machine} 
The atomic structure of glasses and their surfaces depends sensitively on thermal history, requiring melt-quench simulations with long trajectories and large system sizes to adequately sample disorder and relaxation.\cite{vollmayr1996cooling,yu2021structural} For this reason, most prior studies of a-\alumina have relied on classical force fields.\cite{gutierrezMolecularDynamicsStudy2002,adigaAtomisticSimulationsAmorphous2006,carruzzo2021distribution,zhangDependenceGlassTransition2024, lee2024atomistic} While these enable large-scale simulations, their accuracy is fundamentally limited by simplified functional forms and parameterizations optimized for bulk crystals rather than highly defective, glassy surface environments.\cite{matsuiTransferableInteratomicPotential1994} Conversely, \textit{ab initio} molecular dynamics based on density functional theory (DFT) provides a reliable description of complex bonding,\cite{car1985unified} but its steep computational cost restricts accessible system sizes and quench times, precluding statistically converged studies of amorphous surfaces.\cite{car1988structural,sarnthein1995structural,lizarragaStructuralCharacterizationAmorphous2011,sivaramanMachineLearnedInteratomic2020,Harper2023ChemSci,davis2011structural,zhang2021first} Recently, machine-learning interatomic potentials (MLIPs) have emerged as a compelling alternative. By retaining near-first-principles accuracy while achieving orders-of-magnitude speedups, MLIPs enable the large-scale, statistically converged simulations necessary for amorphous materials that bridge the gap between classical force fields and quantum methods.\cite{behlerFourGenerationsHighDimensional2021,deringerMachineLearningInteratomic2019,Xu2021PCCP,Cheng2024PrecisChem, erhard2022machine}

Despite growing interest in MLIPs for disordered materials, a detailed molecular-level understanding of a-\alumina free surfaces remains limited. Existing MLIP studies on alumina are largely restricted either to crystalline phases\cite{rodriguezMartinezWyckoffSitesAlumina2024,zhangExploringEnergyLandscape2025,yangResolvingAtomicStructure2025,duRevealingMolecularStructures2024} or to purely bulk amorphous environments.\cite{liEffectsDensityComposition2020,gramatteUnveilingHydrogenChemical2025} While these amorphous studies have provided valuable insights into density, off-stoichiometry, and hydrogen impurity states in the bulk, they do not address intrinsic glass formation kinetics or free surface phenomena. 
Consequently, atomistic investigations attempting to resolve structural gradients, coordination heterogeneity, and defect statistics at amorphous surfaces has remained tethered almost exclusively to classical force fields,\cite{adigaAtomisticSimulationsAmorphous2006} lacking chemically accurate \textit{ab initio} descriptions. 
Moreover, it remains unclear whether the pronounced structural disorder at amorphous oxide surfaces is accompanied by fundamentally different dynamical behavior compared to the bulk glass. 
In this work, we address these open questions by employing Deep Potential (DP)\cite{zhangDeepPotentialMolecular2018} molecular dynamics to model the complete evolution of a-\alumina from the high-temperature liquid to the vitrified free surface. By first validating the model against experimental liquid structures and the bulk glass transition, we establish a robust foundation for surface investigation. We then use large-scale melt-quench simulations to generate statistically converged amorphous films, enabling us to systematically map structural gradients and identify specific under-coordinated motifs at the free surface. Finally, we evaluate the accompanying dynamical behavior, directly comparing the surface relaxation and glass transition to the bulk interior. Ultimately, this approach delivers a coherent molecular picture of amorphous alumina surfaces grounded in first-principles accuracy.

\section{Results}

\subsection{Structure of liquid \alumina}\label{subsec:liquid}

\begin{figure*}[htbp]
    \centering
    \includegraphics[width=0.9\linewidth]{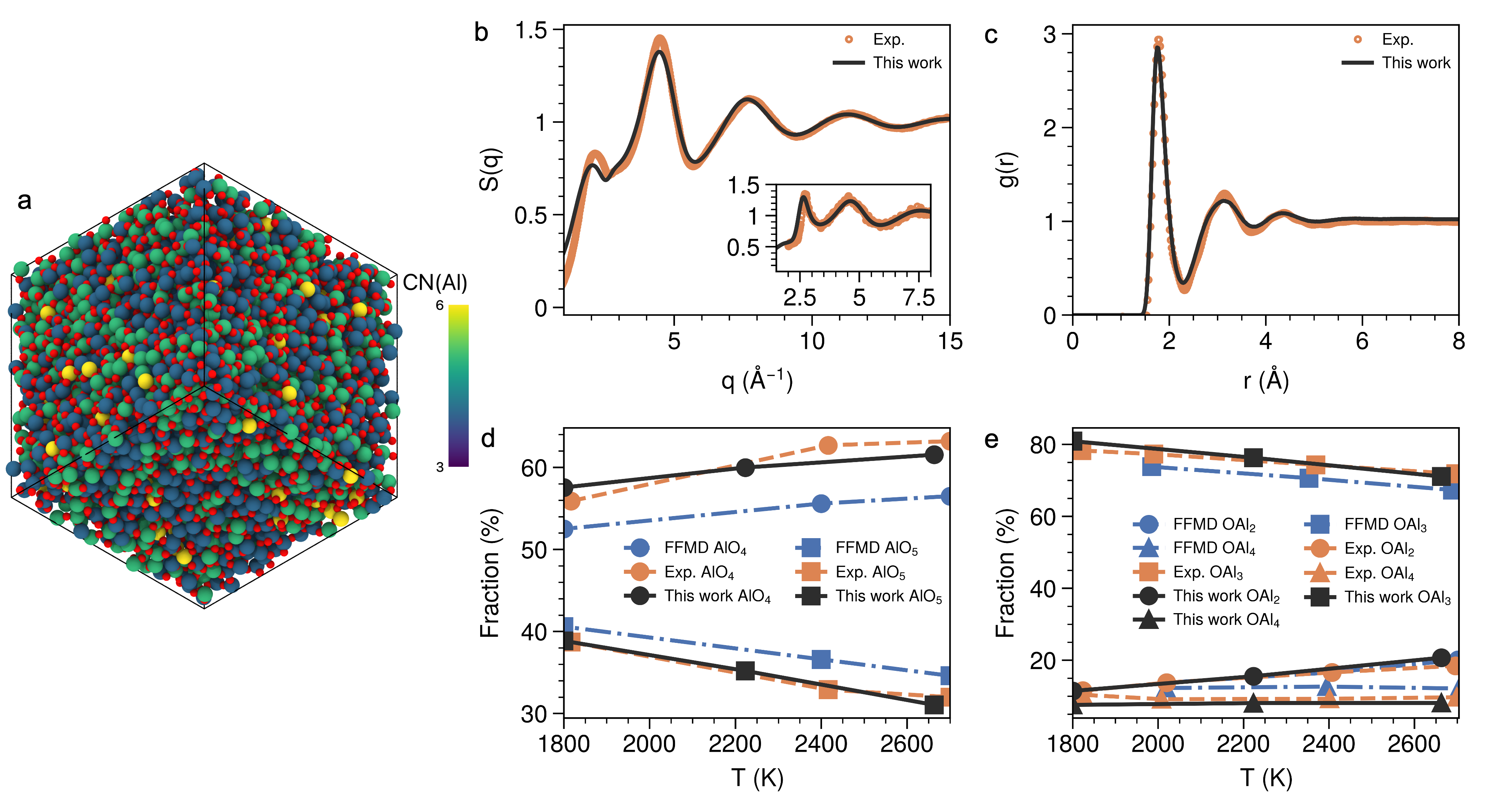}
    \caption{Structural characterization of liquid Al$_2$O$_3$ compared to experimental data from Ref. \cite{shiStructureAmorphousDeeply2019}. (a) Representative equilibrium snapshot of the 10,000-atom simulation cell at 2700~K. O atoms are shown as small red spheres and Al atoms as larger spheres colored by their coordination number (CN). (b) Total X-ray structure factor $S(q)$ at 2700~K. The inset shows the computed neutron scattering structure factor in comparison with experimental data (2587~K). (c) Total X-ray radial distribution function $g(r)$ at 2700~K. (d) Temperature-dependent fractions of Al species with varying CNs. (e) Temperature-dependent fractions of O species with varying CNs. Panels (d,e) include classical force-field results (FFMD) for comparison.\cite{shiStructureAmorphousDeeply2019} The DP model reproduces the experimental  coordination populations noticeably better than the classical force field. }
    \label{fig:str}
\end{figure*}

We first establish the fidelity of the actively trained DP model for bulk \alumina by validating its description of the high-temperature liquid, a prerequisite for physically meaningful melt-quench simulations of glass formation. A 10,000-atom liquid system was simulated over a range of temperatures (computational details are provided in Sec.~\ref{sec:compu}). A representative equilibrium configuration at 2700 K is shown in Fig.~\ref{fig:str}A, illustrating the disordered network of AlO$_n$ polyhedra and the heterogeneous distribution of local coordination environments.

An accurate representation of two-body structural correlations in the melt is essential, as both short- and intermediate-range orders are inherited by the resulting glass. To quantitatively assess the DP model, we compute the total X-ray structure factor $S(q)$ and the radial distribution function $g(r)$ at 2700 K. As shown in Fig.~\ref{fig:str}B, the DP predictions (black solid lines) exhibit excellent agreement with experimental high-energy X-ray and neutron diffraction data (inset) obtained from aerodynamically levitated melts (orange circles),\cite{shiStructureAmorphousDeeply2019} reproducing the positions and relative intensities of the principal diffraction peak and subsequent oscillations. Minor deviations at small $q$ likely arise from finite-size effects, as long-wavelength density fluctuations are inherently limited by the simulation cell dimensions. The $g(r)$ in Fig.~\ref{fig:str}C further confirms that the short-range bonding geometry is faithfully captured, with accurate peak positions and amplitudes for the first Al-O coordination shell and subsequent neighbor shells. The simultaneous agreement in reciprocal- and real-space observables demonstrates that the model reliably captures both local bonding and intermediate-range structural correlations in the liquid state. The corresponding partial structure factors and partial radial distribution functions (including their temperature variation) are provided in the Sec. S2 of the Supporting Information (SI).

Beyond pair correlations, the distribution and thermal evolution of coordination environments provide a more stringent test of the model's chemical realism. Figures~\ref{fig:str}D and \ref{fig:str}E show the temperature-dependent fractions of Al and O coordination species. The DP model reproduces the populations of AlO$_4$ (Al atoms coordinated by four oxygen atoms), AlO$_5$, and OAl$_3$ motifs obtained from empirical potential structure refinement (EPSR) of experimental diffraction data and captures their systematic variation with temperature,\cite{shiStructureAmorphousDeeply2019} reflecting the thermally driven redistribution of bonding environments. In comparison, empirical force-field molecular dynamics (FFMD) exhibits noticeable deviations in coordination fractions.\cite{shiStructureAmorphousDeeply2019} The consistent agreement across pair correlations, coordination statistics, and thermodynamic conditions establishes that the DP model provides an accurate and transferable description of liquid \alumina, forming a robust foundation for investigating bulk glass transition behavior and surface structural heterogeneity.

\subsection{Glass Transition in Bulk Amorphous \alumina}\label{subsec:glass}

\begin{figure}[htbp]
    \centering
    \includegraphics[width=1.0\linewidth]{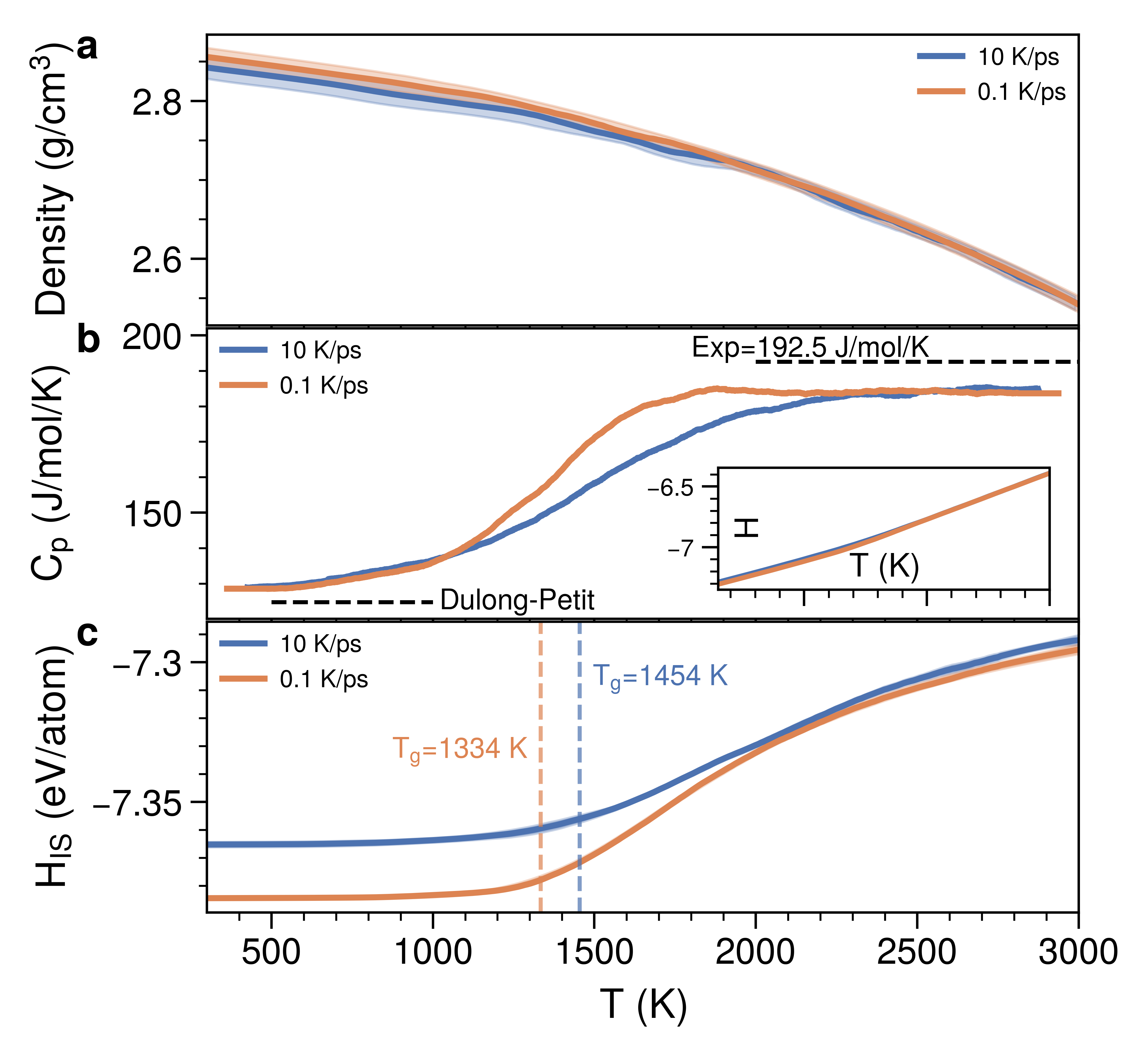}
    \caption{Glass transition of liquid \alumina from Deep Potential melt--quench simulations at cooling rates of 10 and 0.1~K~ps$^{-1}$ (blue and orange, respectively).
    (a) Temperature-dependent mass density $\rho(T)$.
    (b) Isobaric heat capacity $C_p(T)$ obtained by numerical differentiation of the enthalpy $H(T)$ (inset). The computed liquid-state heat capacity at high temperature is in good agreement with the experimental value.
    (c) Inherent-structure enthalpy $H_{\mathrm{IS}}(T)$ determined by energy-minimizing configurations sampled along the cooling trajectory. The glass-transition temperature $T_g$ is identified as the temperature at which $\left|\mathrm{d}^2 H_{\mathrm{IS}}/\mathrm{d}T^2\right|$ is maximized.
    All curves are averaged over four independent cooling replicas; shaded bands in (a) and (c) denote $\pm 1$ standard deviation.}
    \label{fig:glass_transition}
\end{figure}

Having established the accuracy of the DP model for the equilibrium liquid, we now examine whether it captures the non-equilibrium glass transition, which requires both an accurate potential energy surface and simulation timescales far beyond the reach of direct \emph{ab initio} molecular dynamics (AIMD).
Quenching a 10{,}000-atom system from 3000~K to 300~K at our slowest cooling rate (0.1~K~ps$^{-1}$) requires a continuous 27~ns trajectory, whereas AIMD melt-quench studies of \alumina are typically limited to a few hundred atoms and trajectories of tens of picoseconds.\cite{Harper2023ChemSci}
A machine-learned potential of \emph{ab initio} quality is therefore essential for exploring cooling-rate effects on glass formation with statistical confidence.
Figure~\ref{fig:glass_transition}a shows the temperature-dependent mass density during cooling at 10 and 0.1~K~ps$^{-1}$ (blue and orange, respectively).
In both cases, the room-temperature glass density falls within the experimentally measured range, which reflects the strong sensitivity of glass density to preparation conditions: ALD-grown films span 2.7--3.07~g~cm$^{-3}$ depending on deposition temperature,\cite{gorham2014density} while fully dense alumina glass from electrochemical anodization reaches 3.05~g~cm$^{-3}$.\cite{hashimotoStructureAluminaGlass2022} 
In our simulations, slower cooling produces a denser, more relaxed glass, consistent with the expectation that extended liquid-state relaxation allows the system to settle into more efficiently packed configurations.
Figure~\ref{fig:glass_transition}b presents the isobaric heat capacity obtained by differentiating the enthalpy (inset) with respect to temperature.
The computed liquid-state heat capacity at high temperature agrees well with the experimental value of 192.5~J~mol$^{-1}$~K$^{-1}$,\cite{chase1998nist} providing further validation of the DP energy surface in the ergodic regime.
Notably, the limiting values of $C_p$ at both high and low temperatures are essentially independent of cooling rate: in the liquid the system is ergodic and fully samples configuration space regardless of the quench schedule, while in the deep glass the heat capacity is dominated by harmonic vibrations within the inherent-structure basin and approaches the same Dulong--Petit limit ($3R$ per mole of atoms) as the crystal, independent of the particular glassy minimum occupied.
Cooling-rate dependence appears only in the transition region, where the temperature at which the system falls out of equilibrium, and consequently the position and shape of the $C_p$ step, shifts to lower temperatures with slower cooling.

While the density and heat capacity confirm that the DP model produces physically realistic glasses, neither observable provides a sharp marker of $T_g$: the density changes slope gradually across the transition (Fig.~\ref{fig:glass_transition}a), and locating the midpoint of the $C_p$ step requires baseline extrapolations that introduce ambiguity.
A more direct, parameter-free route is offered by the inherent-structure formalism.\cite{stillingerComputerSimulationLocal1985,debenedettiSupercooledLiquidsGlass2001}
By quenching instantaneous configurations sampled along the cooling trajectory to their nearest local energy minimum, one removes thermal vibrations and maps the system's evolution onto the underlying potential energy landscape (PEL).
The resulting inherent-structure enthalpy $H_{\mathrm{IS}}(T)$, shown in Fig.~\ref{fig:glass_transition}c, decreases steadily in the liquid regime as the system explores progressively lower-energy basins, and levels off once structural arrest sets in.
The glass-transition temperature is then identified, without any adjustable parameter, as
\begin{equation}
T_g = \arg\max_T \left| \frac{\mathrm{d}^2 H_{\mathrm{IS}}}{\mathrm{d}T^2} \right|,
\end{equation}
yielding $T_g \approx 1454$~K at 10~K~ps$^{-1}$ and $T_g \approx 1334$~K at 0.1~K~ps$^{-1}$.
The 120~K reduction in $T_g$ with a hundredfold decrease in cooling rate is consistent with the kinetic nature of the glass transition: slower cooling extends the temperature window over which the liquid can relax, allowing the system to descend deeper into the PEL before becoming kinetically trapped, as evidenced by the lower $H_{\mathrm{IS}}$ plateau reached at the slower rate. 
A Vogel--Fulcher--Tammann (VFT) fit to the cooling-rate dependence of $T_g$ returns a divergence temperature $T_0 \approx 0$. 
Extrapolating to a typical experimental quench rate of $\sim$10 K min$^{-1}$ yields a $T_g$ of approximately 894 K (Fig. S6 of the SI). 
This value falls within the experimentally measured range of 743--943 K,\cite{heSolgelDerivedAlumina2019,hashimotoStructureAluminaGlass2022} providing validation of the model's kinetic realism.
The structural signatures accompanying the transition, including sharper short-range order features in the glass relative to the high-temperature liquid, are presented in the SI (Fig.~S5).
The calculated total X-ray structure factor of the quenched glass shows good agreement with experimental data (see SI, Fig.~S4), with substantially improved peak positions and amplitudes compared to classical force-field simulations. Residual deviations in the small-$q$ regime are possibly attributable to the density difference between the simulated glass and the experimental sample, arising from the considerably shorter timescale accessible in simulation.
Together, these results confirm that the DP model, by combining \emph{ab initio} accuracy with the efficiency needed for large-scale, long-time simulations inaccessible to AIMD, reliably describes both the equilibrium liquid and the kinetically arrested glassy state of \alumina. The bulk glasses obtained here serve as starting configurations for the free-surface simulations discussed below.

\subsection{Atomic Structure of the Amorphous \alumina Free Surface}

\begin{figure}[htbp]
    \centering
    \includegraphics[width=1\linewidth]{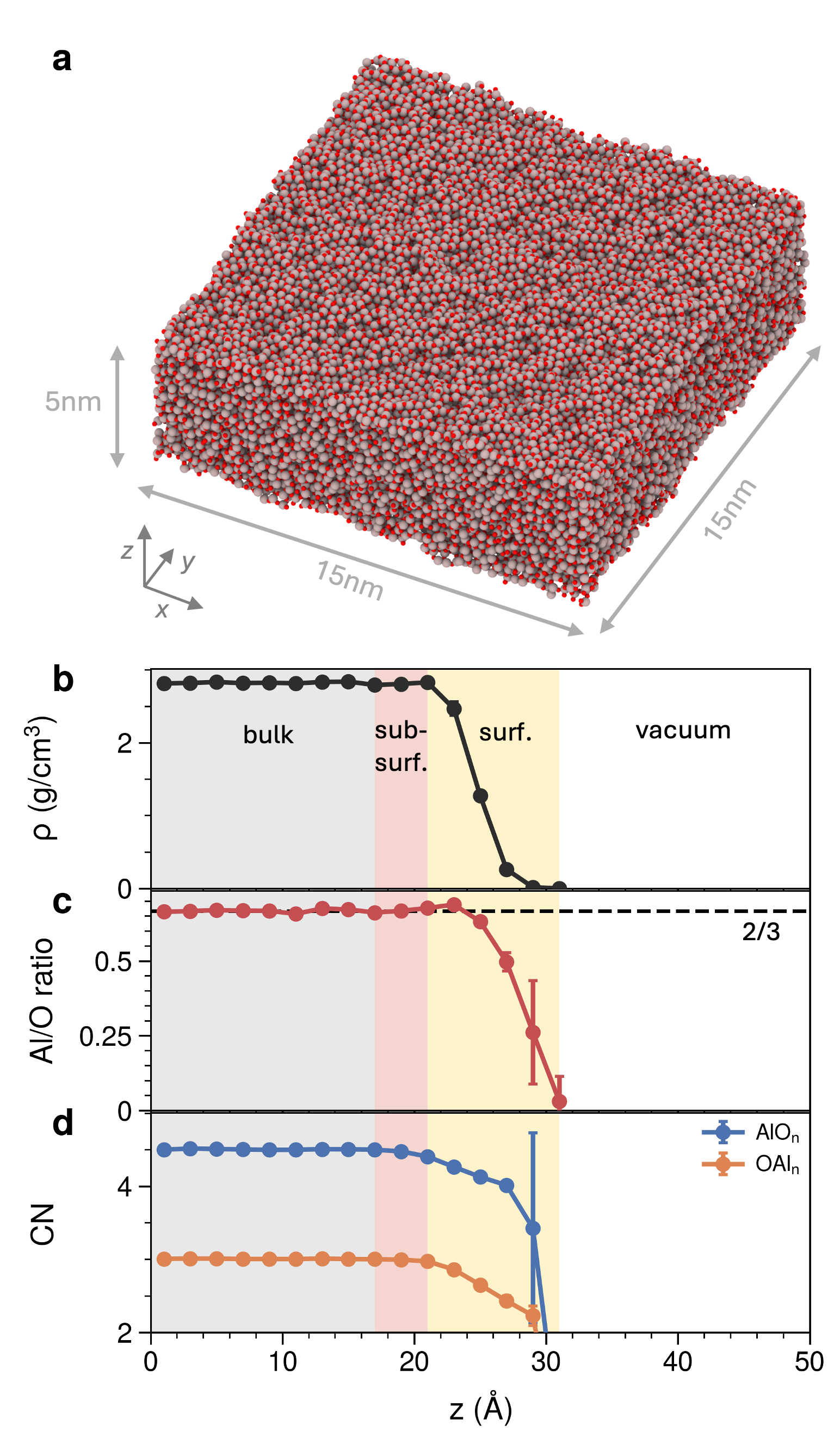}
    \caption{Atomic structure and depth-dependent profiles of a melt-quenched a-\alumina slab with free surfaces. 
    (a) Representative configuration of the $15\times15\times5$~nm$^3$ film (vacuum along $z$); O atoms are shown in red and Al atoms in gray. 
    (b) Laterally averaged mass-density profile $\rho(z)$, highlighting the transition from bulk-like interior to vacuum. 
    (c) Depth-resolved Al/O number-density ratio. The outermost surface is enriched in oxygen (Al/O $<$ 2/3), with a slight compensating aluminum enrichment in the subsurface before the stoichiometric ratio (dashed line) is recovered in the bulk.
    (d) Depth-dependent mean coordination numbers of Al (AlO$_n$) and O (OAl$_n$), showing that coordination recovers more gradually than density across the interfacial region. Shaded regions indicate the bulk, subsurface, and surface zones, defined from the differing convergence depths of the density and coordination profiles. }
    \label{fig:surf_structure}
\end{figure}

To investigate the atomic structure of the a-\alumina free surface, slab models were generated by melt-quenching the largest system considered in this work---an approximate $15\times15\times5$~nm$^3$ film containings 90{,}000 atoms---following the protocol described in Sec.~\ref{sec:compu}.
This system size, far beyond the reach of conventional AIMD, is enabled by the DP model and is essential for capturing the full spatial extent of the disordered surface region while retaining a well-converged bulk-like interior.
A representative quenched configuration is shown in Fig.~\ref{fig:surf_structure}a, where the intrinsic roughness of the amorphous free surface is immediately apparent.
Figure~\ref{fig:surf_structure}b displays the laterally averaged mass-density profile $\rho(z)$: the interior of the film exhibits a constant density consistent with the bulk glass value, followed by a smooth decay to zero across the interfacial region.
The density decay across the interface is well described by a hyperbolic tangent profile,
\begin{equation}
\rho(z) = \frac{\rho_{\mathrm{bulk}}}{2}\left[1 - \tanh\!\left(\frac{z - z_0}{w}\right)\right],
\end{equation}
where $z_0$ is the Gibbs dividing surface position and $w$ characterizes the interfacial width; the depth at which $\rho(z)$ converges to within 1\% of $\rho_{\mathrm{bulk}}$ defines the density-based boundary of the surface region.
The depth-resolved Al/O number-density ratio (Fig.~\ref{fig:surf_structure}c) reveals that the outermost surface is enriched in oxygen, with the ratio falling well below the stoichiometric value of 2/3.
A slight compensating aluminum enrichment appears in the subsurface before the bulk stoichiometry is recovered, consistent with the tendency of oxide surfaces to terminate with the more electronegative species, as previously reported from
classical force-field simulations and inferred experimentally from synchrotron X-ray scattering.\cite{adigaAtomisticSimulationsAmorphous2006,youngProbingAtomicScaleStructure2020}
The mean CNs of both Al and O (Fig.~\ref{fig:surf_structure}d) remain measurably reduced across the interfacial region, recovering fully only deeper into the film interior, more gradually than the mass density.
An analogous sigmoid fit to the mean AlO$_n$ profile yields a deeper convergence boundary, indicating that the Al coordination environment recovers over a longer length scale than the mass density.
This decoupling motivates the schematic division into three zones shown in Fig.~\ref{fig:surf_structure}b: a \emph{surface} region ($\sim$10~\AA{} thick) where the density has not yet reached the bulk value, a \emph{subsurface} region ($\sim$4~\AA{} thick) where the density has recovered but the local coordination environment (particularly the AlO$_n$ distribution) has not, and the \emph{bulk} interior where both are converged.

\begin{figure}[htbp]
    \centering
    \includegraphics[width=0.9\linewidth]{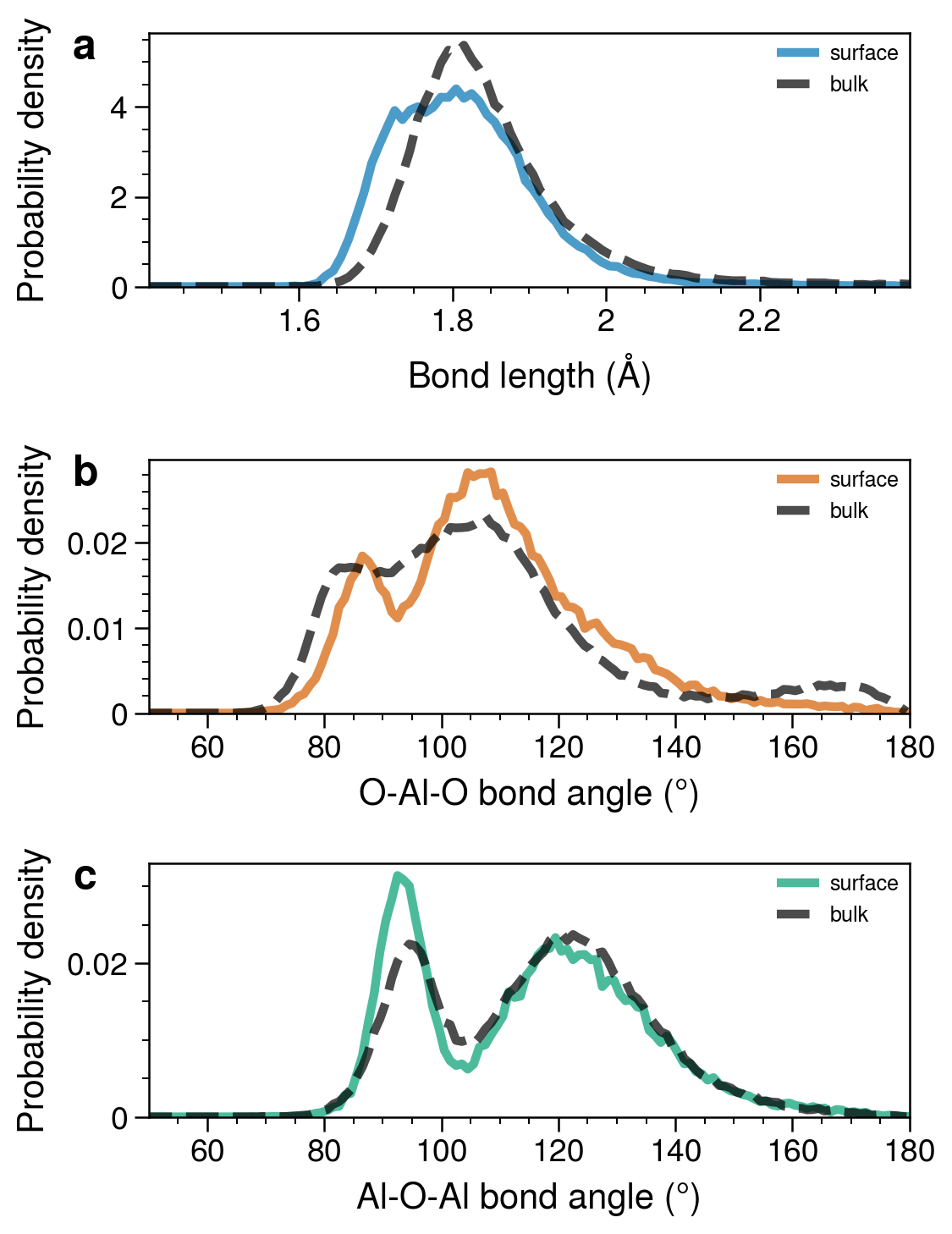}
    \caption{Local bonding geometry in surface versus bulk-like regions of a-\alumina. Probability densities of (a) Al--O bond lengths, (b) O--Al--O bond angles, and (c) Al--O--Al bond angles for atoms classified as interfacial (solid colored lines) or bulk-like interior (gray dashed lines) using the intrinsic-interface definition described in Sec.~\ref{sec:compu}.}
    \label{fig:surf_bond_angle_distributions}
\end{figure}

Beyond the depth-resolved profiles, the local bonding geometry at the surface differs systematically from the bulk interior.
Figure~\ref{fig:surf_bond_angle_distributions}a compares the Al--O bond-length distributions for atoms classified as surface or bulk-like using the intrinsic-interface definition described in Sec.~\ref{sec:compu}.
The surface distribution is broader and shifted toward shorter bond lengths, consistent with the prevalence of under-coordinated Al sites: fewer surrounding oxygen neighbors allow the remaining Al--O bonds to contract, a trend well established in molecular orbital and crystal-chemistry arguments.\cite{adigaAtomisticSimulationsAmorphous2006,gutierrezMolecularDynamicsStudy2002}
The intra-polyhedral O--Al--O bond-angle distribution (Fig.~\ref{fig:surf_bond_angle_distributions}b) likewise reflects the coordination change. In the bulk, the distribution is bimodal, with a peak near 80--90$^{\circ}$ from cis-edges of AlO$_5$ and AlO$_6$ polyhedra and a second peak near $\sim$109$^{\circ}$ characteristic of AlO$_4$ tetrahedra. At the surface the $\sim$90$^{\circ}$ shoulder is suppressed and the tetrahedral peak dominates, consistent with the increased fraction of AlO$_4$ and AlO$_3$ sites.
The inter-polyhedral Al--O--Al bridging-angle distribution (Fig.~\ref{fig:surf_bond_angle_distributions}c) shows an enhanced population near $\sim$90$^{\circ}$ at the surface, suggesting increased edge-sharing polyhedral connectivity compared to the bulk network.
This increase in edge-sharing, despite the overall reduction in CN, is consistent with the oxygen-enriched surface identified above: with fewer bridging oxygens available, adjacent polyhedra share edges more extensively to maintain network connectivity.
Collectively, these changes show that the amorphous free surface is not simply a truncated bulk but develops a distinct bonding character shaped by the reduced coordination environment.

\subsection{Coordination Environments at the a-\alumina Free Surface}

\begin{figure*}[htbp]
    \centering
    \includegraphics[width=1\linewidth]{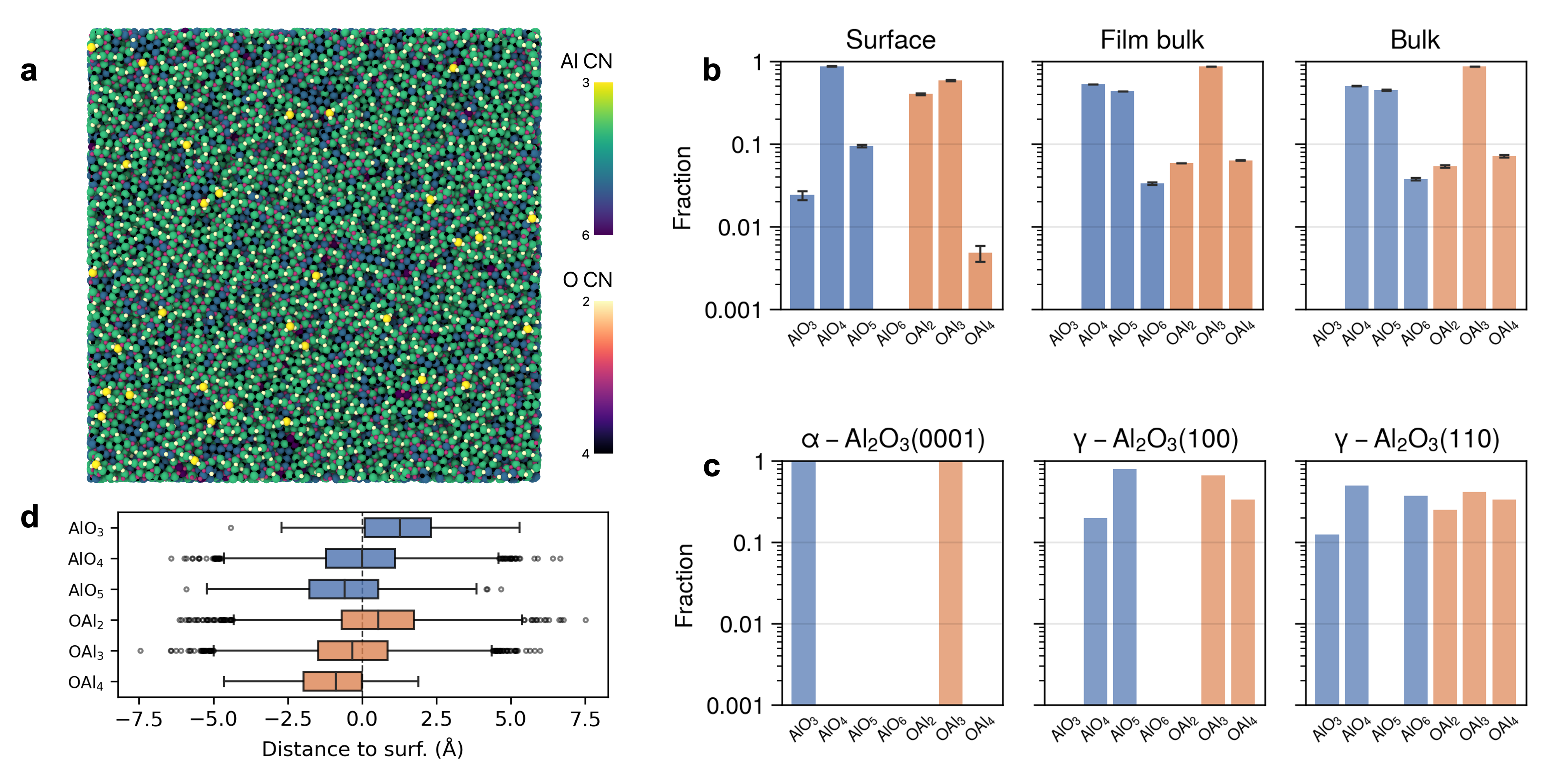}
    \caption{Coordination environments at the a-\alumina surface
        and comparison to bulk and crystalline references.
        (a) Top-down view of the surface layer showing topologically
        heterogeneous spatial distribution of Al and O coordination environments.
        Al atoms (larger spheres) and O atoms (smaller spheres) are colored by their respective coordination numbers as indicated by the color bars.
        (b) Fractions of AlO$_n$ (blue) and OAl$_n$ (orange) species at the
        amorphous surface, in the film interior (``Film bulk''), and in an
        independent periodic bulk-glass simulation (``Bulk''); the $y$ axis
        is logarithmic to resolve low-population motifs. Error bars denote
        $\pm 1$ standard deviation over four independent replicas.
        (c) Corresponding coordination fractions at $\alpha$-\alumina(0001) and
        $\gamma$-\alumina(100)/(110) unrelaxed terminations, for comparison to the amorphous surface.
        (d) Vertical displacement of each coordination species relative to the mean intrinsic-interface height (Sec.~\ref{sec:compu}). Positive values indicate atoms protruding outward (toward vacuum), while negative values indicate atoms recessed inward (toward the film interior). Box plots show the median, interquartile range, and whiskers extending to 1.5 times the interquartile range; outliers are shown as individual points.}
    \label{fig:surf_coordination}
\end{figure*}

The a-\alumina surface hosts a diverse range of Al and O coordination motifs that are spatially intermixed without mesoscopic clustering or segregation, as shown by the coordination map in Fig.~\ref{fig:surf_coordination}a.
Quantitatively, for atoms classified as surface species using the intrinsic-interface definition (Sec.~\ref{sec:compu}), AlO$_4$ accounts for approximately 88\% of surface Al, followed by AlO$_5$ ($\sim$9.5\%) and the under-coordinated AlO$_3$ ($\sim$2.4\%), with AlO$_6$ contributing $<$0.1\%; the corresponding O speciation is dominated by OAl$_3$ ($\sim$59\%), with OAl$_2$ constituting a substantial $\sim$41\% of surface oxygen and OAl$_4$ remaining below 1\% (Fig.~\ref{fig:surf_coordination}b, left; note the logarithmic $y$-axis used to resolve the minority species).
The two sublattices respond asymmetrically to surface truncation: OAl$_2$, a minority species in the bulk glass ($\sim$11\%), rises to $\sim$41\% at the surface, while AlO$_3$ jumps from $\lesssim$0.5\% in the bulk to $\sim$2.4\%---indicating that oxygen under-coordination is amplified at the surface while aluminum under-coordination is created by it.
The coordination distributions of the film interior and an independent periodic bulk-glass simulation (Fig.~\ref{fig:surf_coordination}b, center and right) are nearly identical, with AlO$_4$/AlO$_5$/OAl$_3$ fractions agreeing to within 1\% (absolute) across the two systems, confirming that these modifications are genuinely surface-specific.
In contrast to the broad continuous distributions at the amorphous surface, the crystalline alumina facets shown in Fig.~\ref{fig:surf_coordination}c exhibit narrow, discrete coordination spectra dictated by their crystallographic termination. The Al-terminated $\alpha$-\alumina(0001) surface consists exclusively of AlO$_3$ and OAl$_3$ species, while $\gamma$-\alumina(100) (proposed by Digne et al.\cite{Digne2004JCatal}) is dominated by AlO$_4$ and OAl$_3$, and $\gamma$-\alumina(110) retains predominantly AlO$_6$ and OAl$_4$ motifs reflecting its more bulk-like cation environment. We emphasize that these crystalline panels are intended only to illustrate the qualitative contrast between the discrete coordination spectra of ordered facets and the broad continuum sampled by the amorphous surface; they are not a rigorous comparison of surface energetics, and the specific terminations shown correspond to unrelaxed cleavage planes rather than the lowest-energy reconstructions of each facet.
Alternative $\gamma$-\alumina models (e.g., the distorted-anion-lattice structure of Yang, Shang, and Liu\cite{yangResolvingAtomicStructure2025}) would yield shifted coordination fractions on these facets but would not change the qualitative contrast between discrete crystalline spectra and the continuous amorphous distribution emphasized here.
Overall, the amorphous surface, lacking any preferred termination plane,  samples a broad continuum of coordination environments that encompasses and extends beyond those found on any single crystalline facet.

Beyond the coordination census, the spatial organization of these motifs reveals additional structure.
Figure~\ref{fig:surf_coordination}d shows that under-coordinated species (AlO$_3$, OAl$_2$) preferentially occupy topographic protrusions above the mean intrinsic surface, while higher-coordinated motifs (AlO$_5$, OAl$_4$) are recessed into valleys, establishing a direct link between nanoscale surface roughness and the local coordination environment.
The same trend is reflected in the laterally averaged $z$-resolved coordination profiles (Fig.~S7), where the bulk-dominant AlO$_5$ and OAl$_3$ species are progressively replaced toward the free surface by AlO$_4$, AlO$_3$, and OAl$_2$, with OAl$_2$ ultimately becoming nearly as populous as OAl$_3$ at the outermost ridges---a striking inversion of the OAl$_3$-dominated bulk distribution.
The reduction of Al and O coordination is moreover spatially correlated: AlO$_3$ sites are preferentially bonded to OAl$_2$ oxygens rather than to the OAl$_3$ species that dominate the bulk (Fig.~S8), indicating that when the polyhedral network is truncated, the missing connectivity affects both the Al and its neighboring O atoms simultaneously.
The prevalence of OAl$_2$ at the outermost surface further implies a local reduction in network cross-linking, as each such oxygen bridges only two Al polyhedra rather than three.
Connectivity analysis shows that approximately 40\% of surface OAl$_2$ species are isolated, while the remaining $\sim$60\% participate in short connected segments---predominantly pairs and triplets of OAl$_2$ units sharing Al neighbors---that decay rapidly in length, with no extended chain-like motifs beyond five to six units (Fig.~S9).

\begin{figure}[htbp]
    \centering
    \includegraphics[width=1.0\linewidth]{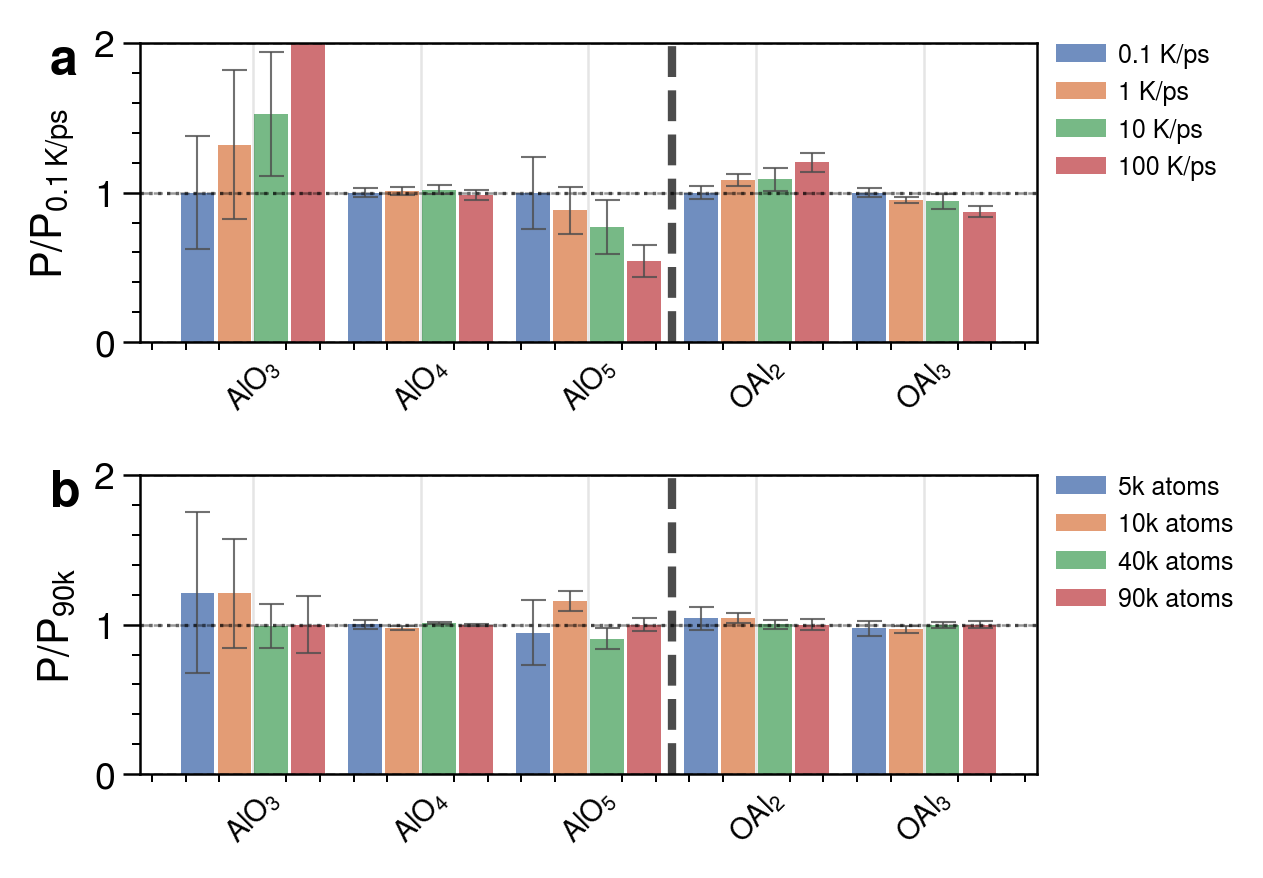}
    \caption{Dependence of surface coordination environments on cooling rate and system size, shown as ratios relative to a reference. Surface atoms are identified using the intrinsic-interface definition (Sec.~\ref{sec:compu}). $P$ denotes the fraction of surface atoms in each motif.
    (a) Ratios $P / P_{0.1\,\mathrm{K/ps}}$ for AlO$_n$ and OAl$_n$ species at four cooling rates for a 10{,}000-atom system, normalized to the slowest cooling rate. Faster cooling increases the populations of under-coordinated motifs (AlO$_3$, OAl$_2$), consistent with a less relaxed glass retaining greater structural heterogeneity.
    (b) Ratios $P / P_{90\mathrm{k}}$ for four system sizes with fixed film thickness, at a fixed cooling rate of 1~K~ps$^{-1}$, normalized to the largest system. The dotted horizontal line at unity marks the reference; the dashed vertical line separates Al-centered from O-centered motifs.}
    \label{fig:CN_surf_effect}
\end{figure}

The surface coordination distributions are sensitive to thermal history but insensitive to system size.
Figure~\ref{fig:CN_surf_effect}a shows the surface AlO$_n$ and OAl$_n$ populations across four cooling rates relative to the slowest one ($P/P_{0.1\,\mathrm{K/ps}}$).
Faster cooling progressively enriches the under-coordinated motifs AlO$_3$ and OAl$_2$ and depletes the higher-coordinated AlO$_5$ and OAl$_3$ environments.
This is the coordination-level manifestation of the glass-transition physics discussed in the preceding subsection: a higher effective quench temperature traps the system in a shallower basin of the PEL, preserving a broader distribution of less ordered local environments.
Conversely, slower cooling yields a better-annealed surface depleted in under-coordinated sites, implying that the population of reactive low-CN sites at amorphous alumina surfaces can, in principle, be tuned through thermal processing.
In contrast, system size has no systematic effect on the same populations across more than an order of magnitude (Fig.~\ref{fig:CN_surf_effect}b), with relative deviations from the largest system well below the cooling-rate-induced changes. Smaller systems do show larger statistical uncertainty for low-abundance motifs owing to the reduced number of surface atoms, but the point estimates remain consistent with the larger systems.
This size-independence is robust, as an alternative analysis defining the surface by its Cartesian depth confirms that the coordination fractions remain generally invariant across these scales (see Fig.~S10 in the SI).
Therefore, thermal history, specifically the relaxation timescale set by the cooling rate, rather than system size is the primary handle on surface structural heterogeneity.

\subsection{Surface versus Bulk: Glass Transition and Relaxation Dynamics}

The glass transition at the surface is not only of fundamental interest but also governs how thermal history affects the stability and aging of amorphous films, as well as the accessibility of surface sites for adsorption and reaction.
In the bulk, this transition is commonly probed by density $\rho(T)$ or the inherent-structure enthalpy $H_{\mathrm{IS}}(T)$. However, at a rough, heterogeneous surface, these thermodynamic observables do not have well-defined, locally resolved analogs.
We therefore use the temperature dependence of the mean CNs during cooling as a topological proxy for kinetic arrest: a crossover from liquid-like (coordination changing with $T$) to glass-like (coordination frozen) behavior would signal a surface glass transition.

\begin{figure}[htbp]
    \centering
    \includegraphics[width=0.9\linewidth]{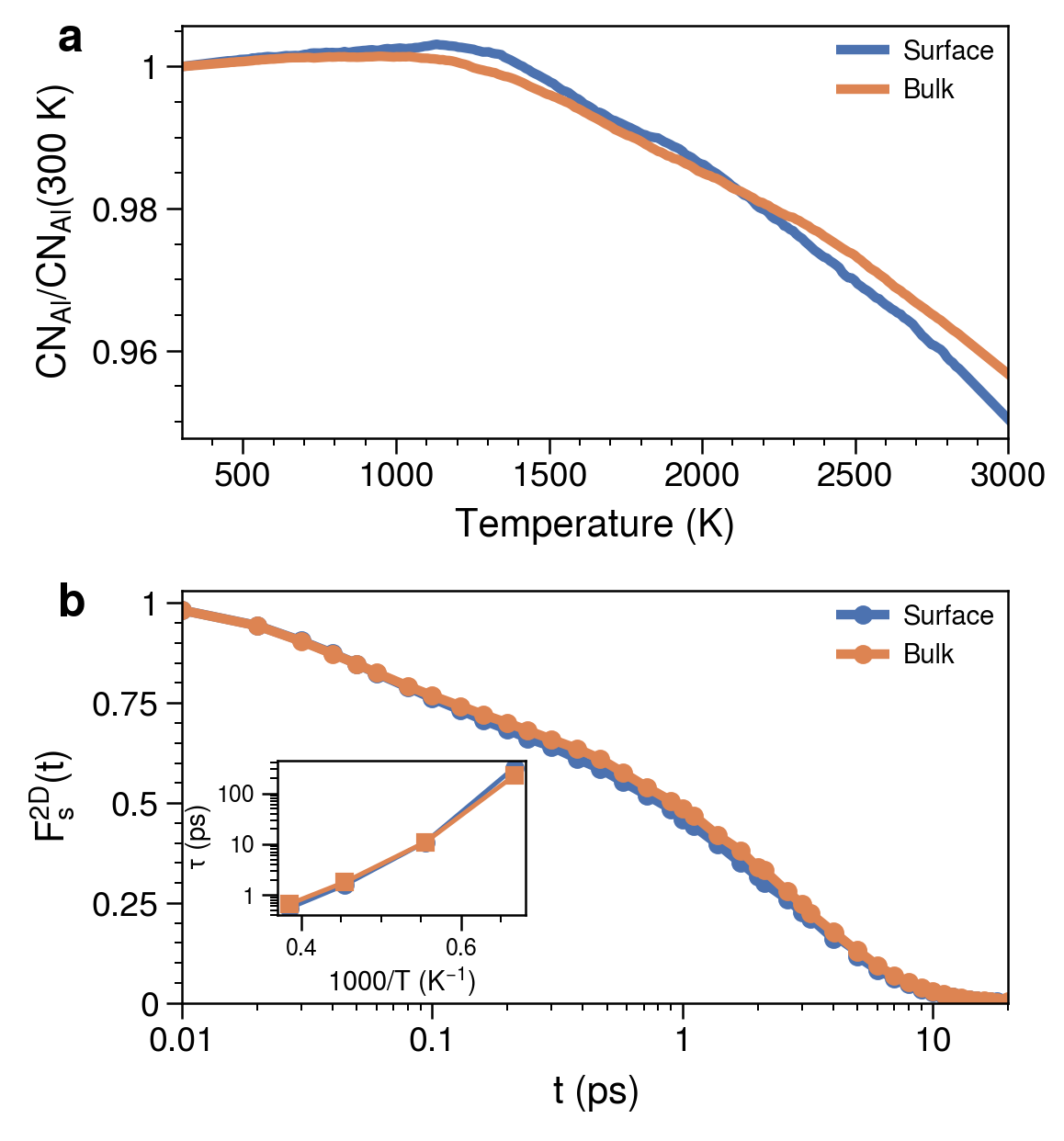}
    \caption{Comparison of surface and bulk dynamics. (a) Temperature dependence of the mean Al coordination number, normalized by its 300~K value, for surface and bulk-like regions during cooling from 3000~K to 300~K with a fixed rate of 0.1 K~ps$^{-1}$. (b) Two-dimensional self-intermediate scattering function $F_s^{2\mathrm{D}}(t)$, restricted to in-plane ($x$-$y$) displacements, for atoms in the surface and bulk-like regions at a representative high temperature (2200~K), showing comparable structural relaxation. The inset shows the relaxation time $\tau$ as a function of inverse temperature for both regions.}
    \label{fig:surf_diffusion}
\end{figure}

Figure~\ref{fig:surf_diffusion}a shows the mean CN of Al atoms, normalized by its value at 300~K, for surface and bulk atoms during cooling from 3000~K to 300~K at a rate of 0.1~K/ps. 
In the bulk, the coordination evolution exhibits a crossover at $\sim$1300~K, consistent with $T_g$ determined from the inherent-structure enthalpy (Fig.~\ref{fig:glass_transition}c). 
The surface displays a qualitatively similar crossover at a comparable temperature, indicating that the surface and bulk vitrify at nearly the same $T_g$ despite their pronounced structural differences.
Above $T_g$, the mean surface coordination changes more steeply with temperature than its bulk counterpart, reflecting the lower average connectivity at the surface: with fewer constraints per atom, the local topology can reorganize more readily until vitrification freezes it in place, after which the coordination plateaus as in the bulk.

The near-coincidence of the surface and bulk glass-transition temperatures is a nontrivial result, since enhanced surface mobility that is widely reported for molecular and metallic glasses would be expected to lower the surface $T_g$ relative to the bulk. 
To probe relaxation dynamics more directly, we compute the two-dimensional self-intermediate scattering function $F_s^{2\mathrm{D}}(q,t)$ restricted to the lateral $x$-$y$ plane,
\begin{equation}
    F_s^{2\mathrm{D}}(q,t) = \frac{1}{N} \left\langle \sum_{j=1}^{N} \exp\!\bigl[i\,\mathbf{q}_{\parallel}\cdot\bigl(\mathbf{r}_{j,\parallel}(t)-\mathbf{r}_{j,\parallel}(0)\bigr)\bigr] \right\rangle
\end{equation}
where $\mathbf{q}_{\parallel}$ is a wavevector lying in the surface plane, $\mathbf{r}_{j,\parallel}$ denotes the in-plane position of atom $j$, and $|\mathbf{q}_{\parallel}|$ is chosen to correspond to the principal peak of the two-dimensional structure factor. 
The restriction to in-plane displacements avoids the ambiguity of the normal component for surface atoms. 
For the isotropic bulk, $F_s^{2\mathrm{D}}(t)$ yields relaxation times consistent with the standard three-dimensional $F_s(t)$, so the same in-plane definition can be applied to surface and bulk regions on equal footing.
Figure~\ref{fig:surf_diffusion}b presents $F_s^{2\mathrm{D}}(t)$ for surface and bulk atoms at a representative supercooled liquid temperature (2200~K). 
The two decay curves are nearly identical, and the extracted relaxation times $\tau$ (defined by $F_s^{2\mathrm{D}}(\tau)=1/e$; inset) agree across the temperature range examined.
Together, the coordination evolution and dynamical analyses establish that the a-\alumina surface, despite its markedly different local structure, vitrifies at a comparable $T_g$ and relaxes on timescales comparable to the bulk interior.

\section{Discussion}

A natural question, recently raised in the literature,\cite{gramatte2024we} is whether MLIPs are genuinely necessary for amorphous metal oxides, or whether classical force fields (possibly recalibrated against existing \textit{ab initio} databases) already suffice. The present work argues that, for a-\alumina and especially for its surfaces, the answer is unambiguously yes, and the reasoning rests on three observations. First, the bulk liquid and glass results presented in Sec.~\ref{subsec:liquid} and Sec.~\ref{subsec:glass} show that DPMD reproduces experimental pair-correlation functions, structure factors, and coordination statistics noticeably better than the widely used classical force fields, demonstrating that \textit{ab initio}-quality interactions are required to capture even the bulk structure faithfully. Second, although AIMD would in principle provide the same accuracy, its cost and poor scaling with system size make it infeasible to simulate either the slow melt-quench protocols needed to produce well-relaxed glasses or the large slab geometries needed to resolve surface heterogeneity with adequate sampling. DPMD resolves this tension by combining \textit{ab initio} accuracy with classical-MD-like efficiency, enabling the 90{,}000-atom slabs, multiple independently quenched replicas, and cumulative trajectory times exceeding $10^2$~ns required for the present study. 
Third, and crucial for surface work specifically, the active-learning protocol incorporated liquid, glassy, \textit{and} free-surface configurations, so that the model is not extrapolated to environments absent from its training data---a flexibility that classical force fields, fit to a fixed set of bulk reference structures, fundamentally lack. 
The combination of accuracy, efficiency, and training-set flexibility is what makes the present surface analysis possible, and these capabilities are directly transferable to other amorphous oxides and to disordered interfaces more broadly.

The practical impact of these methodological advantages becomes clear when comparing our predicted surface structures with the only prior atomistic study of a-\alumina free surfaces, the classical force-field simulations of Adiga \textit{et al.}\cite{adigaAtomisticSimulationsAmorphous2006}
Both studies observe the same qualitative features: oxygen enrichment at the outermost surface, a compensating aluminum enrichment slightly below, and reduced mean Al coordination at the surface relative to the interior. However, several quantitative differences emerge. 
First, the structural transition zone in the classical simulations extends only $\sim$4~\AA{} from the surface, compared to $\sim$10~\AA{} in the present work (Fig.~\ref{fig:surf_structure}).
Because the two studies employ comparable system sizes ($\sim$21{,}000 versus 90{,}000 atoms), this discrepancy is unlikely to be a finite-size effect; instead, it suggests that the Matsui potential might produce an artificially smooth surface with insufficient structural disorder and roughness. 
Second, the depth-resolved compositional and coordination profiles of Adiga \textit{et al.} exhibit large oscillations arising from the use of a single quenched configuration, whereas the present profiles, averaged over multiple independently quenched replicas, are statistically well converged. 
Third, the Matsui potential overestimates AlO$_4$ and underestimates AlO$_5$ populations in the bulk: 78\% vs. 20\%, compared to 52\% vs. 44\% from our DPMD (see SI, Fig.~S7) and experimental EPSR values of $\sim$50\% vs. $\sim$42\%.\cite{shiStructureAmorphousDeeply2019} 
This bulk bias propagates to the surface coordination distribution, altering the predicted populations of reactive under-coordinated sites. 
Beyond these methodological contrasts, the surface coordination distributions obtained here reveal structural and chemical features that have not been characterized in earlier work, which we discuss next.

The broad coordination census at the amorphous surface carries direct implications for surface chemistry. The under-coordinated Al sites (AlO$_3$, together with the surface-enriched AlO$_4$) constitute Lewis acid centers, while the oxygen-enriched outermost layer, characterized by a high OAl$_2$ population and reduced cross-linking, provides complementary Br\o nsted base character. 
Unlike crystalline facets with discrete coordination spectra (Fig.~\ref{fig:surf_coordination}c), the amorphous surface samples a continuous distribution of environments. This generates a correspondingly continuous spectrum of acid--base site strengths, likely underlying the broad reactivity distributions reported experimentally on amorphous alumina supports.\cite{mavricAdvancedApplicationsAmorphous2019} Furthermore, these low-CN Al and O sites are spatially correlated: OAl$_2$ oxygens preferentially neighbor AlO$_3$ (and vice versa). This is consistent with bond-valence (Pauling) compensation, as the lowest-CN Al provides the strongest individual Al-O bond to offset the oxygen's missing third neighbor, implying that Lewis acid and Br\o nsted base sites are actively co-localized. Finally, OAl$_2$ sites do not aggregate into extended chains (SI, Fig.~S9) but occur predominantly as isolated centers or short clusters. Thus, reactive base sites are dispersed across the surface rather than concentrated in patches. These bare-surface motifs establish the precise structural starting point for future reactive simulations of water dissociation and atomic layer deposition (ALD), where the spatial distribution of low-CN sites will dictate the initial hydroxyl coverage and the balance of bridging versus terminal OH groups.

Another notable finding is that the amorphous surface and bulk interior exhibit comparable glass-transition temperatures and structural-relaxation times (Fig.~\ref{fig:surf_diffusion}), in contrast to the enhanced surface mobility widely reported for organic and metallic glasses.
This absence of a mobile surface layer likely reflects the strongly directional, partially covalent character of the Al--O network, which suppresses the cooperative rearrangements that lower the surface $T_g$ in systems with weaker or more isotropic bonding. 
More generally, the magnitude of surface mobility enhancement has been shown to correlate with glass fragility,\cite{li2022surface} with strong network formers exhibiting only modest surface acceleration; amorphous SiO$_2$, the prototypical strong glass, likewise shows surface dynamics nearly indistinguishable from the bulk. The bulk-like surface dynamics observed here for a-\alumina are therefore consistent with this broader picture, placing it toward the strong-glass regime as far as surface dynamics is concerned.
This has practical implication for thin-film growth.
In molecular and metallic glasses, physical-vapor-deposited films can achieve exceptional stability, so-called ultrastable glasses, because enhanced surface mobility allows each newly arriving layer to relax toward the supercooled-liquid equilibrium before being buried by subsequent deposition.\cite{swallen2007organic} 
This mechanism requires a fast surface layer,\cite{berthier2017origin,annamareddy2021mechanisms} which the present results show is absent in a-\alumina: the surface relaxes on essentially the same timescale as the bulk.
Consequently, deposited a-\alumina films cannot exploit the ultrastable-glass route, and the surface coordination motifs identified here in melt-quenched models are expected to be representative of those in ALD, sputtered, or e-beam-evaporated films grown at comparable effective quench rates, rather than dramatically less relaxed than them. The structural heterogeneity characterized in this work should therefore also reflect deposited films, not be an artifact of the melt-quench protocol.

Several limitations of the present study should be noted. First, the slowest cooling rate explored here (0.1 K ps$^{-1}$) remains orders of magnitude faster than typical experimental deposition rates; the systematic cooling-rate dependence presented in Fig.~\ref{fig:CN_surf_effect}a partially mitigates this by providing a controlled basis for extrapolating coordination populations toward slower, more experimentally relevant effective quench rates. Second, all surfaces studied here are pristine and anhydrous, whereas real amorphous alumina surfaces are invariably hydroxylated under ambient conditions. The bare-surface coordination census and spatial-motif analysis presented here are intended precisely as the structural input from which future reactive simulations of water dissociation, hydroxylation, and ALD precursor chemistry can be launched. Extending the present DP model to wet interfaces will require additional active-learning iterations on hydroxylated and hydrated surface configurations, which were deliberately excluded from the present training set in order to keep the scope focused on bare-surface structure. Finally, the PBE functional used to generate the training data lacks an explicit treatment of dispersion interactions; while this has minimal impact on the structural and dynamical observables reported here, it should be revisited when the potential is applied to molecular adsorption energetics, where vdW contributions can be significant.


\section{Conclusion}

In this work, we have employed Deep Potential molecular dynamics to construct large-scale, \textit{ab initio}-quality models of amorphous \alumina bulk glasses and melt-quenched free surfaces, enabling a systematic characterization of surface structure, coordination environments, and dynamics with a statistical confidence inaccessible to prior methods. The DP model quantitatively reproduces experimental liquid and glass pair-correlation functions, structure factors, and coordination statistics, and captures the cooling-rate dependence of the bulk glass transition, demonstrating that machine-learned potentials are not merely a substitute for classical force fields but a necessary tool for amorphous oxides where existing empirical models systematically misrepresent polyhedral populations. At the free surface, we identify a finite structural transition region in which the mass density recovers more rapidly than the local coordination, an oxygen-enriched outermost layer with reduced network cross-linking, and a broad distribution of under-coordinated Al and O motifs whose populations are governed by glass stability. The under-coordinated Al and O sites are not randomly distributed but spatially paired, with OAl$_2$ oxygens preferentially neighboring AlO$_3$ in a manner consistent with bond-valence compensation, while OAl$_2$ sites themselves remain dispersed rather than aggregating into extended chains. Despite this pronounced structural heterogeneity, the surface exhibits a glass-transition temperature and structural-relaxation dynamics comparable to those of the bulk interior, indicating that the disordered surface structure, once formed, is kinetically stable. 
These findings establish a molecular-level structural foundation for the acid--base, hydroxylation, and precursor-binding chemistry of amorphous alumina interfaces relevant to catalysis, atomic layer deposition, and thin-film technologies. The bare-surface coordination map provided here is the natural starting point for reactive simulations of dissociative water adsorption, and, longer term, for extension to amorphous oxide heterointerfaces and to enhanced-sampling techniques capable of reaching experimentally relevant thermal histories.

\section{Methodology}
\subsection{Machine learning interatomic potential}

To enable large-scale simulations of amorphous Al$_2$O$_3$ with near \emph{ab initio} accuracy, we employed the Deep Potential Molecular Dynamics (DPMD) framework implemented in the DeePMD-kit package,\cite{Wang2018CPC,zeng2023JCP} which represents the potential energy surface using a deep neural network trained on first-principles data. In this approach, the total energy of the system is decomposed as
\begin{equation}
E = \sum_{i=1}^{N} E_i(\mathcal{D}_i)
\end{equation}
where $E_i$ is the atomic energy contribution predicted by the neural network from the descriptor $\mathcal{D}_i$ encoding the local chemical environment of atom $i$ within a finite cutoff. This decomposition ensures translational, rotational, and permutational invariance by construction. DPMD has been shown to accurately reproduce first-principles energetics and forces while achieving orders-of-magnitude computational speedups compared to \emph{ab initio} molecular dynamics, making it well suited for melt-quench simulations and statistically converged modeling of amorphous materials.
The local chemical environment is defined by neighboring atoms within a smooth cutoff radius of 6~\AA, with a switching function applied between an inner cutoff of 0.5~\AA\ and the outer cutoff to ensure continuous forces.
The ``se\_e2\_a'' descriptor is adopted to represent the local atomic environment with an embedding network of $24 \times 48 \times 96$.
The dimension of the axis neuron is set to 16.
The fitting network from the descriptor to the atomic energy is $256 \times 256 \times 256$.
The model parameters were optimized by minimizing the loss function
\begin{equation}
\mathcal{L} = p_e \, \Delta E^2 + \frac{p_f}{3N} \sum_{i=1}^{N} \| \mathbf{F}_{i}^{\mathrm{DP}} - \mathbf{F}_{i}^{\mathrm{DFT}} \|^2
\end{equation}
where $\Delta E = (E^{\mathrm{DP}} - E^{\mathrm{DFT}})/N$ is the per-atom energy error and $\mathbf{F}_i$ denotes the force on atom $i$. The prefactors $p_e$ $p_f$ were adjusted during training following the standard schedule: $p_e$ was ramped from 0.02 to 1, $p_f$ from 100 to 1 over 200 epochs, with the learning rate decaying exponentially from $1\times10^{-3}$ to $1\times10^{-8}$.

The training dataset was generated through an iterative active-learning procedure implemented in the GDPy package designed to efficiently sample relevant regions of configuration space.\cite{Xu2026GDPy} 
Initial configurations were drawn from crystalline, liquid, and amorphous Al$_2$O$_3$ structures, including snapshots along melt-quench trajectories. At each iteration, four DPMD models were trained independently and used to perform molecular dynamics simulations; the model deviation for each atom $i$ was estimated as
\begin{equation}
\sigma_f^{i} = \sqrt{\frac{1}{M} \sum_{m=1}^{M} \| \mathbf{F}_i^{(m)} - \bar{\mathbf{F}}_i \|^2},
\end{equation}
where $M=4$ is the number of models and $\bar{\mathbf{F}}_i$ is the ensemble-averaged force.\cite{Zhang2019PRM}
Configurations with $\max_i \sigma_f^{i}$ falling between the lower and upper thresholds $\sigma_f^{\mathrm{lo}}$ and $\sigma_f^{\mathrm{hi}}$ (set to 0.08 and 0.24~eV/\AA, respectively) were selected as candidates for new DFT labeling, thereby enriching the dataset with structurally and thermodynamically diverse environments while avoiding redundant or unphysical configurations. This cycle was repeated until the model predictions converged and no significant extrapolation was detected during production simulations of both bulk and surface systems.

All reference energies and forces used for training were computed using DFT as implemented in the Vienna \emph{Ab initio} Simulation Package (VASP).\cite{kresseEfficiencyAbinitioTotal1996} 
The Perdew-Burke-Ernzerhof (PBE) exchange–correlation functional within the generalized gradient approximation was employed, together with projector augmented-wave (PAW) pseudopotentials provided in the VASP distribution. 
A plane-wave kinetic energy cutoff of 400~eV was used throughout. Brillouin-zone sampling was $\Gamma$-centered with a spacing of 0.04 \r{A}$^{-1}$ in each direction of the reciprocal space.\cite{Monkhorst1976PRB}

The trained model achieves a root mean squared error (RMSE) of 0.124 eV/{\AA} for atomic forces, and 0.003 eV/atom for total energies. 
The dataset comprises 11,629 structures in total, and the details on the composition of the training set are provided in Sec. S1 of the Supporting Information (SI).

\subsection{Computational details}\label{sec:compu}

DPMD simulations were performed in LAMMPS using a velocity--Verlet integration scheme with a 1~fs timestep.\cite{LAMMPS}
Periodic boundary conditions were applied in all three dimensions.
Temperature and pressure are controlled using a Nos\'{e}-Hoover thermostat and barostat with coupling time constants of 100 fs and 1000 fs, respectively.\cite{martyna1992nose} 
All NPT simulations were carried out at a target pressure of 1~bar.

\paragraph{Bulk amorphous \alumina.} 
Bulk liquid and glassy alumina systems were simulated in a cubic cell containing 10,000 atoms (2000 \alumina formula units).
Initial configurations were generated by melting structures obtained from short preparatory simulations using the Matsui classical force field;\cite{matsuiTransferableInteratomicPotential1994} these served solely to provide starting coordinates, and all reported properties were obtained from subsequent DP trajectories.
Bulk simulations were conducted in the NPT ensemble for both liquid equilibration and melt-quench processes, allowing the system density to adapt naturally during glass formation.
The system was equilibrated at target liquid temperatures (e.g. 2700 K) for 5~ns, followed by production runs of 5~ns.
Equilibration of the liquid state was confirmed by convergence of the potential energy and density, as well as by sufficiently large mean-squared displacements (see Sec.~S2 of the SI).
\alumina glasses were generated by quenching equilibrated liquid configurations from 3000~K to 300~K using a constant cooling rate.
Three cooling rates, 0.1, 1, and 10~K~ps$^{-1}$, were employed to examine thermal-history effects on the resulting glass structure.

\paragraph{Amorphous \alumina free surface.}
a-\alumina films with free surfaces were generated following procedures analogous to the bulk melt-quench protocol from 3000 K to 300 K.
Initial slab configurations were constructed from bulk glassy structures by introducing vacuum layers of 5~nm thickness along the surface-normal ($z$) direction; this vacuum thickness is sufficient to eliminate spurious slab--slab interactions through the periodic images, as verified by the convergence of the density profile to zero well within the vacuum region.
The lateral ($x$-$y$) cell dimensions were fixed to the equilibrium values obtained from the NPT bulk simulation at 300~K for the corresponding cooling rate.
Surface simulations were performed in the NVT ensemble, as the presence of vacuum allows the slab density to relax naturally without the need for external pressure control.
Five lateral system sizes were considered to assess finite-size effects, with total atom counts ranging from approximately 2500 to 90,000 atoms.
The largest system corresponds to a slab of dimensions 15~nm~$\times$~15~nm~$\times$~5~nm (excluding the vacuum region).
Unless otherwise specified, results presented in the main text are obtained from this largest system.
To improve statistical sampling, at least four independently quenched configurations were analyzed for both bulk and surface systems.
These configurations were generated using independent DP models trained with different random seeds on the same underlying DFT dataset.

\paragraph{Identification of surface atoms.} Surface atoms were identified using an intrinsic-interface construction based on a coarse-grained (Gaussian-smeared) atomic number density field following the Willard--Chandler approach.\cite{willardInstantaneousLiquidInterfaces2010}
Specifically, an instantaneous density field was defined as
\begin{equation}
\tilde{\rho}(\mathbf{r},t)=\sum_{i=1}^{N}\left(2\pi\sigma^{2}\right)^{-3/2}
\exp\!\left[-\frac{\left|\mathbf{r}-\mathbf{r}_{i}(t)\right|^{2}}{2\sigma^{2}}\right],
\end{equation}
where $\sigma$ is the coarse-graining width (set to 1.5~\AA\ in this work) and $\mathbf{r}_{i}$ are atomic coordinates.
The instantaneous free surface was then defined as the isosurface $\tilde{\rho}(\mathbf{r},t)=\rho^{\ast}$, with $\rho^{\ast}$ chosen as half the bulk number density (i.e.\ $\rho^{\ast} = \rho_{\mathrm{bulk}}/2$).
Atoms were classified as surface atoms if their shortest distance to this isosurface was smaller than a cutoff $d_{\mathrm{s}} = 1$~\AA.
The same procedure was applied to both sides of the slab, and all reported surface-resolved properties were averaged over the two sides and over independently quenched configurations.

\paragraph{Structural analysis.}
Partial radial distribution functions $g_{\alpha\beta}(r)$ and partial structure factors $S_{\alpha\beta}(q)$ were computed from the simulation trajectories using standard definitions for a multicomponent system.
Total scattering functions for comparison with X-ray and neutron diffraction data were obtained using the Faber--Ziman formalism,\cite{faber1965theory} in which the total structure factor is
\begin{equation}
S(q) = 1 + \sum_{\alpha,\beta} w_{\alpha\beta} \bigl[ S_{\alpha\beta}(q) - 1 \bigr],
\end{equation}
with weights
\begin{equation}
w_{\alpha\beta} = \frac{(2-\delta_{\alpha\beta})\, c_\alpha\, c_\beta\, b_\alpha\, b_\beta}{\left(\sum_\gamma c_\gamma\, b_\gamma\right)^2},
\end{equation}
where $c_\alpha$ is the atomic fraction of species $\alpha$, $\delta_{\alpha\beta}$ is the Kronecker delta, and $b_\alpha$ denotes the scattering weight of species $\alpha$.
For X-ray scattering, $b_\alpha$ was approximated by the $q \to 0$ atomic form factors ($f_{\mathrm{Al}} = 13.26$, $f_{\mathrm{O}} = 8.00$), neglecting their gradual decay with $q$. Over the range shown in Fig.~\ref{fig:str}, this approximation only weakly affects the total $S(q)$ because the dominant Al-O Faber--Ziman weight remains nearly unchanged.
For neutron scattering, the coherent scattering lengths $b_{\mathrm{Al}} = 3.45$~fm and $b_{\mathrm{O}} = 5.80$~fm were used.
The total radial distribution function $g(r)$ was obtained analogously by weighting the partial $g_{\alpha\beta}(r)$ with the same $w_{\alpha\beta}$ coefficients.

Coordination numbers (CNs) were determined by counting the number of Al--O pairs within a cutoff distance of $r_{\mathrm{cut}} = 2.5$~\AA, corresponding to the first minimum of the partial Al--O radial distribution function $g_{\mathrm{AlO}}(r)$.
This cutoff is consistent with values adopted in previous experimental and computational studies of amorphous and liquid alumina.\cite{leeStructureAmorphousAluminum2009,skinner_joint_2013,hashimotoStructureAluminaGlass2022}
For each Al atom, the CN is defined as the number of O neighbors within $r_{\mathrm{cut}}$, yielding the AlO$_n$ ($n = 3$--6) species reported throughout this work; analogously, for each O atom the CN counts the number of Al neighbors, yielding OAl$_n$ ($n = 1$--4) species.

\section*{Acknowledgment}
This work was conducted within the ``Chemistry in Solution and at Interfaces'' (CSI) Center funded by the USA Department of Energy under Award DE-SC0019394.
The work was also funded by Shell International Exploration and Production Inc., USA. 
Simulations and analyses were performed using resources from Princeton Research Computing at Princeton University, which is a consortium led by the Princeton Institute for Computational Science and Engineering (PICSciE) and Office of Information Technology’s Research Computing. These resources include a GPU-based computing cluster purchased with support from the National Science Foundation (Grant No. NSF-MRI: OAC-2320649).

\section*{Data Availability}
The trained Deep Potential model, the active-learning training dataset, and representative amorphous \alumina bulk and free-surface configurations generated in this work are openly available at \url{https://github.com/CSIprinceton/amorphous_alumina.git}.

\bibliography{alumina,xjy}

\end{document}


\title[]{Supporting Information for ``From Bulk to Surface: Structure and Dynamics of Amorphous Alumina from Deep Potential Molecular Dynamics''}

\author{Zheng Yu}
\thanks{These authors contribute equally to this work}
\affiliation{Department of Chemistry, Princeton University, Princeton, NJ, 08540, United States}
\author{Jiayan Xu}
\thanks{These authors contribute equally to this work}
\affiliation{Department of Chemistry, Princeton University, Princeton, NJ, 08540, United States}
\author{Abhirup Patra}
\email{Abhirup.Patra@shell.com}
\affiliation{Shell International Exploration \& Production Inc., 200 N Dairy Ashford Rd, Houston, Texas 77079, United States}
\author{Sharan Shetty}
\affiliation{Shell India Markets Pvt., Ltd., Mahadeva Kodigehalli, Bengaluru 562149, Karnataka, India}
\author{Detlef Hohl}
\affiliation{Shell Information Technology International Inc., 3333 Highway 6 South, Houston, Texas 77082, United States}
\author{Roberto Car}
\email{Correspondence: rcar@princeton.edu}
\affiliation{Department of Chemistry, Princeton University, Princeton, NJ, 08540, United States}


\maketitle

\tableofcontents

\newpage

\section{Training dataset and model accuracy}
\begin{figure}[hbtp]
    \centering
    \includegraphics[width=0.80\linewidth]{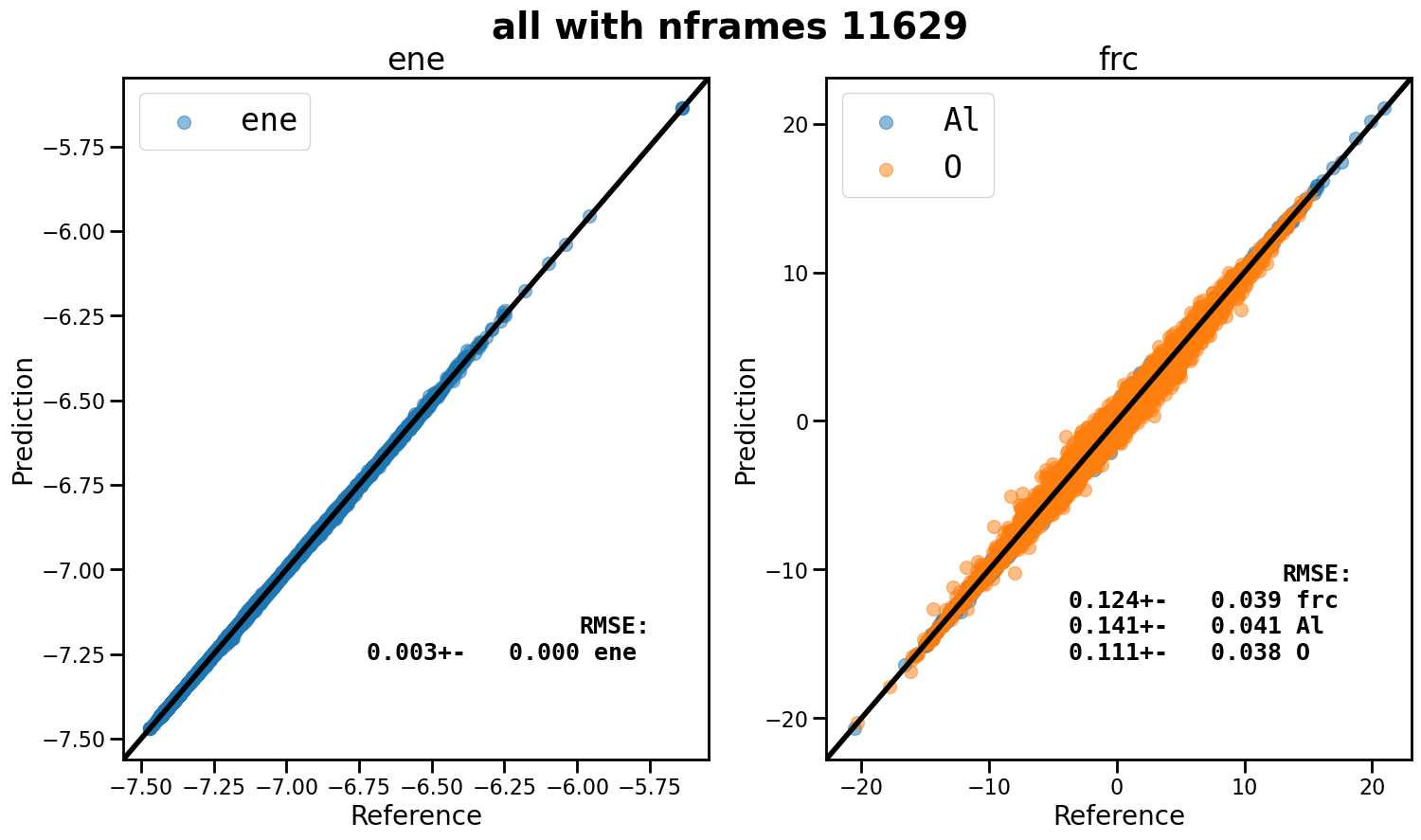}
    \caption{Parity plots of (left) energies and (right) forces comparing DFT reference values and DP model predictions across the full training dataset (11\,629 structures). RMSEs for each subsystem are listed in the inset.}
    \label{fig:model_rmse}
\end{figure}

The structures in the training dataset were generated from DPMD simulations through an active learning workflow.
A timestep of 2 fs was used to improve computational efficiency.
Crystalline bulk and surface structures were sampled using NVT simulations at five temperatures (150, 300, 600, 1200, and 2400 K).
Bulk sampling started from the $\gamma$-Al$_2$O$_3$ structure proposed by Digne.\cite{Digne2004JCatal}
The corresponding unit cell contains 40 atoms.
Crystalline surfaces were cleaved from either $\gamma$-Al$_2$O$_3$ or $\alpha$-Al$_2$O$_3$ bulk structures with different facets.
For the $\gamma$-Al$_2$O$_3$ system, the dataset includes $p(1\times1)$-, $p(2\times1)$-, and $p(3\times2)$-(100) surfaces as well as $p(1\times1)$- and $p(2\times2)$-(110) surfaces.
For the $\alpha$-Al$_2$O$_3$ system, a $p(3\times2\sqrt3)$ surface was considered.
Melt-quenched bulk structures were generated through melt--quench simulations starting from cubic $\alpha$-Al$_2$O$_3$ bulk structures with two sizes: $p(1\times3\times1)$ (180 atoms) and $p(2\times3\times1)$ (360 atoms).
The simulations were performed under NPT conditions at 1 bar following a three-stage protocol: the systems were first equilibrated at 3800 K for 0.5 ns, then cooled from 3800 K to 300 K over 3.5 ns, and finally equilibrated at 300 K for an additional 0.5 ns.
Melt-quenched surface structures were generated by introducing a 20 \AA\ vacuum layer along the $z$-direction to the melt-quenched bulk systems described above.
These surfaces were simulated under NVT conditions using a three-stage temperature protocol: heating from 300 K to 1800 K over 1.5 ns, cooling from 1800 K to 300 K over 1.5 ns, followed by equilibration at 300 K for 0.5 ns.
The number of structures in each subsystem, together with the root-mean-squared errors (RMSEs) of energies and forces, are summarized in Table~\ref{tab:model_rmse}.
Parity plots comparing DFT and DP predictions for energies and forces across the entire dataset are shown in Fig.~\ref{fig:model_rmse}.

\begin{table}[hbtp]
    \centering
    \caption{Number of training structures and root-mean-squared errors (RMSEs) in energy and forces for each subsystem of the training dataset.}
    \label{tab:model_rmse}
    \makebox[\textwidth][c]{
    {
    \begin{tabular}{l c c c}
        \hline
        System &
        Number &
        E$_\mathrm{RMSE}$ [eV] $\downarrow$ &
        F$_\mathrm{RMSE}$ [eV/\AA] $\downarrow$ \\
        \hline
        Crystalline Bulk       &  3932  &         0.001 &         0.048 \\
        Crystalline Surface    &  3266  &         0.003 &         0.070 \\
        Melt-Quenched Bulk     &  3153  &         0.003 &         0.148 \\
        Melt-Quenched Surface  &  1278  &         0.003 &         0.143 \\
        \hline
    \end{tabular}
    }
    }
\end{table}

\section{Liquid alumina}

\begin{figure}[hbtp]
    \centering
    \includegraphics[width=0.8\linewidth]{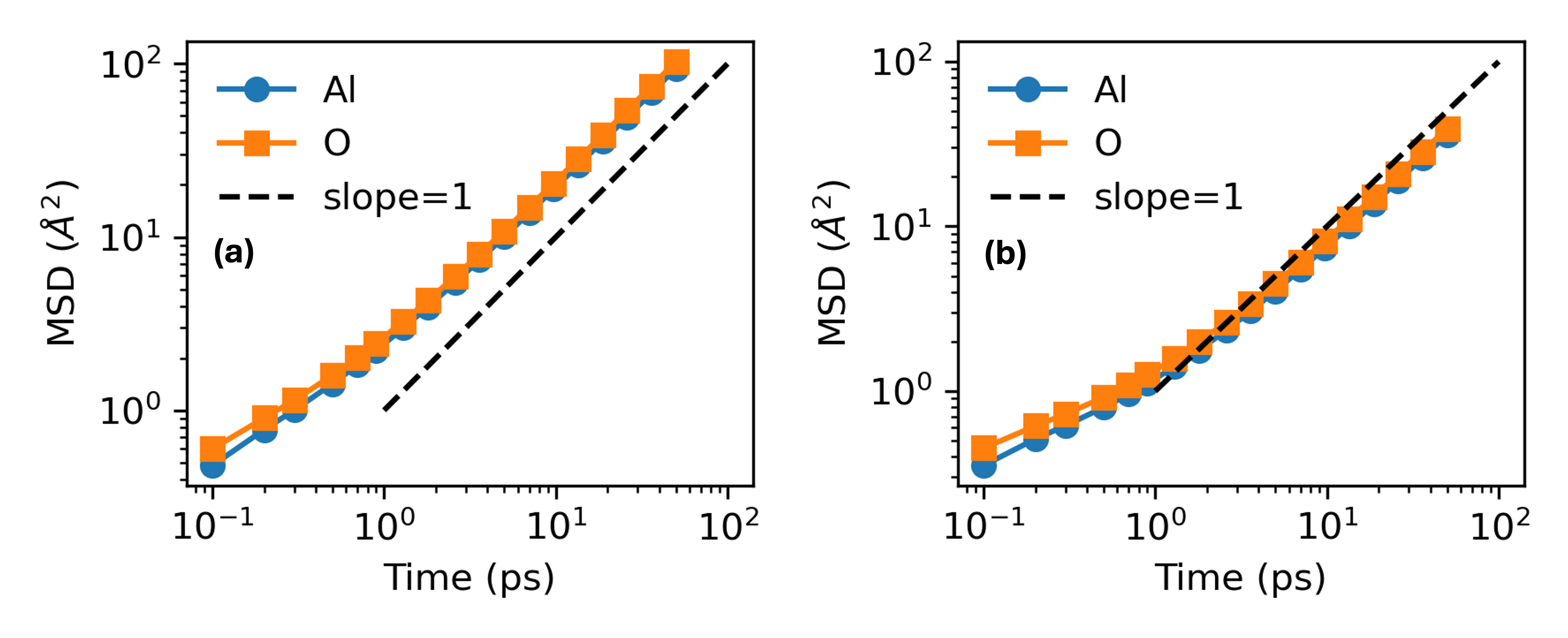}
    \caption{Mean squared displacements (MSD) of liquid \alumina at 2700~K (left) and 2223~K (right). At both temperatures, the MSD enters the Fickian regime within 100~ps, where MSD scales linearly with simulation time.}
    \label{fig:msd_liquid}
\end{figure}

Figure~\ref{fig:msd_liquid} shows the mean squared displacements (MSD) of liquid \alumina at 2700~K and 2223~K. At both temperatures, the MSD enters the Fickian (diffusive) regime within 100~ps, confirming that the production trajectories are long enough to sample equilibrium liquid dynamics.

\begin{figure}[hbtp]
    \centering
    \includegraphics[width=0.75\linewidth]{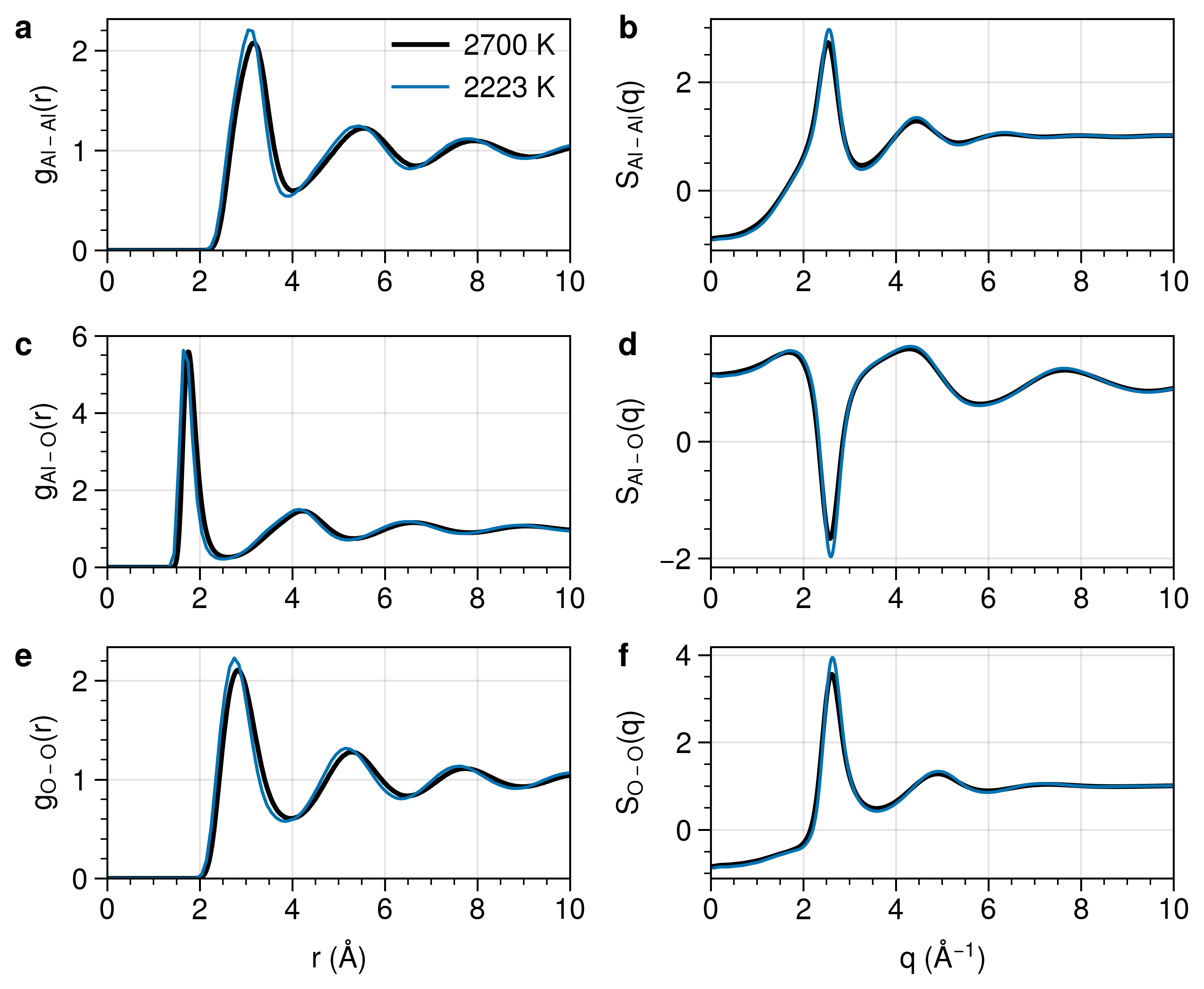}
    \caption{Partial radial distribution functions $g(r)$ and static structure factors $S(q)$ for Al--Al, Al--O, and O--O pairs in liquid \alumina from DPMD simulations at 2700~K and 2223~K.}
    \label{fig:rdf_sq_liquid}
\end{figure}

The partial radial distribution functions (RDFs) $g(r)$ and static structure factors $S(q)$ for Al--Al, Al--O, and O--O pairs at 2700~K and 2223~K are shown in Fig.~\ref{fig:rdf_sq_liquid}. The peak positions and relative intensities are weakly temperature-dependent, indicating that the short-range order in the liquid is robust across this temperature range.

\section{Bulk amorphous alumina}

\begin{figure}[hbtp]
    \centering
    \includegraphics[width=0.6\linewidth]{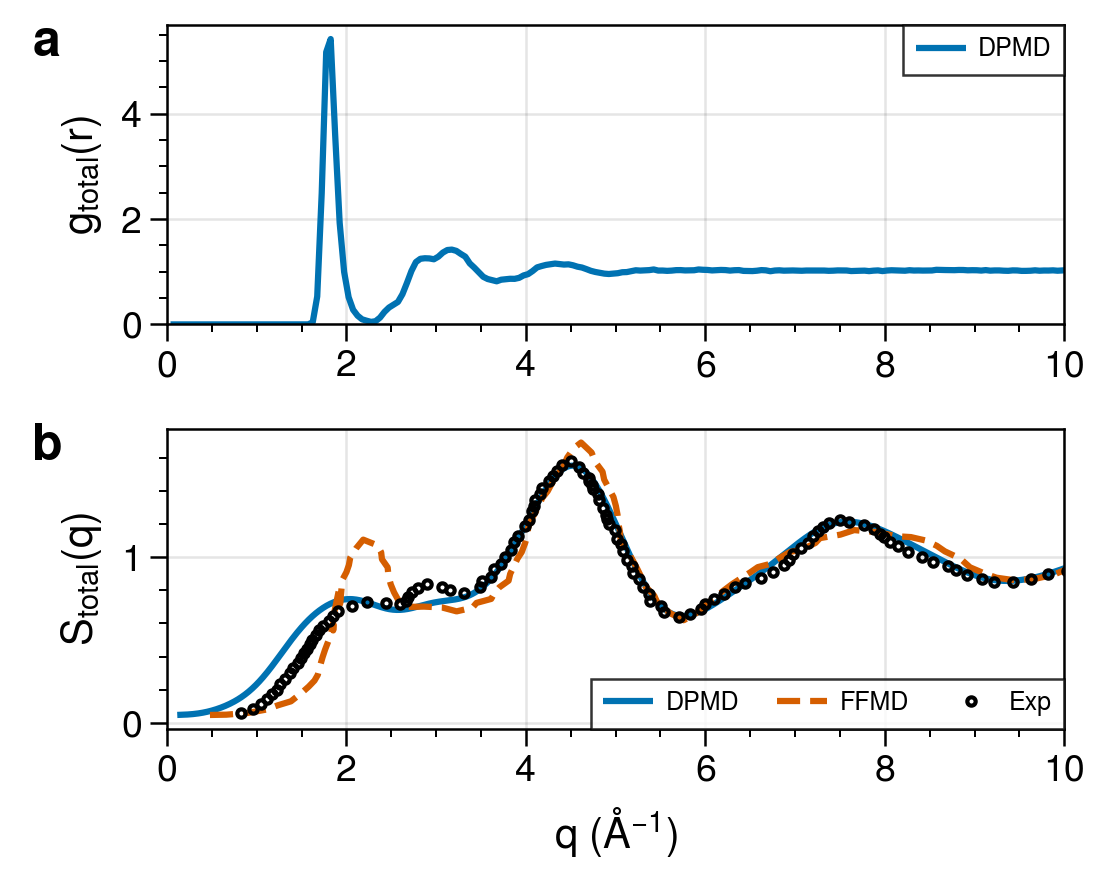}
    \caption{(a) Total radial distribution function $g_\mathrm{total}(r)$ and (b) total X-ray structure factor $S_\mathrm{total}(q)$ of bulk amorphous \alumina from DPMD using a cooling rate of 0.1~K/ps, compared with classical force field results and experimental X-ray scattering data (circles).}
    \label{fig:glass_xray}
\end{figure}

Figure~\ref{fig:glass_xray} shows the total RDF $g_\mathrm{total}(r)$ and the X-ray structure factor $S_\mathrm{total}(q)$ of bulk amorphous \alumina obtained by melt-quenching at 0.1~K/ps.
For comparison, results from classical force fields (FFMD in the figure) are also included, which show notably larger deviations from the experimental data than the DP model.\cite{shiStructureAmorphousDeeply2019} The DP-predicted $S(q)$ is in good agreement with the experimental X-ray scattering data, validating the accuracy of the DP model for the amorphous phase. The discrepancy at $q<2$ {\AA}$^{-1}$ is possibly due to the density difference between the simulated and experimental glasses, arising from the much faster cooling rate used in simulations.

\begin{figure}[hbtp]
    \centering
    \includegraphics[width=\linewidth]{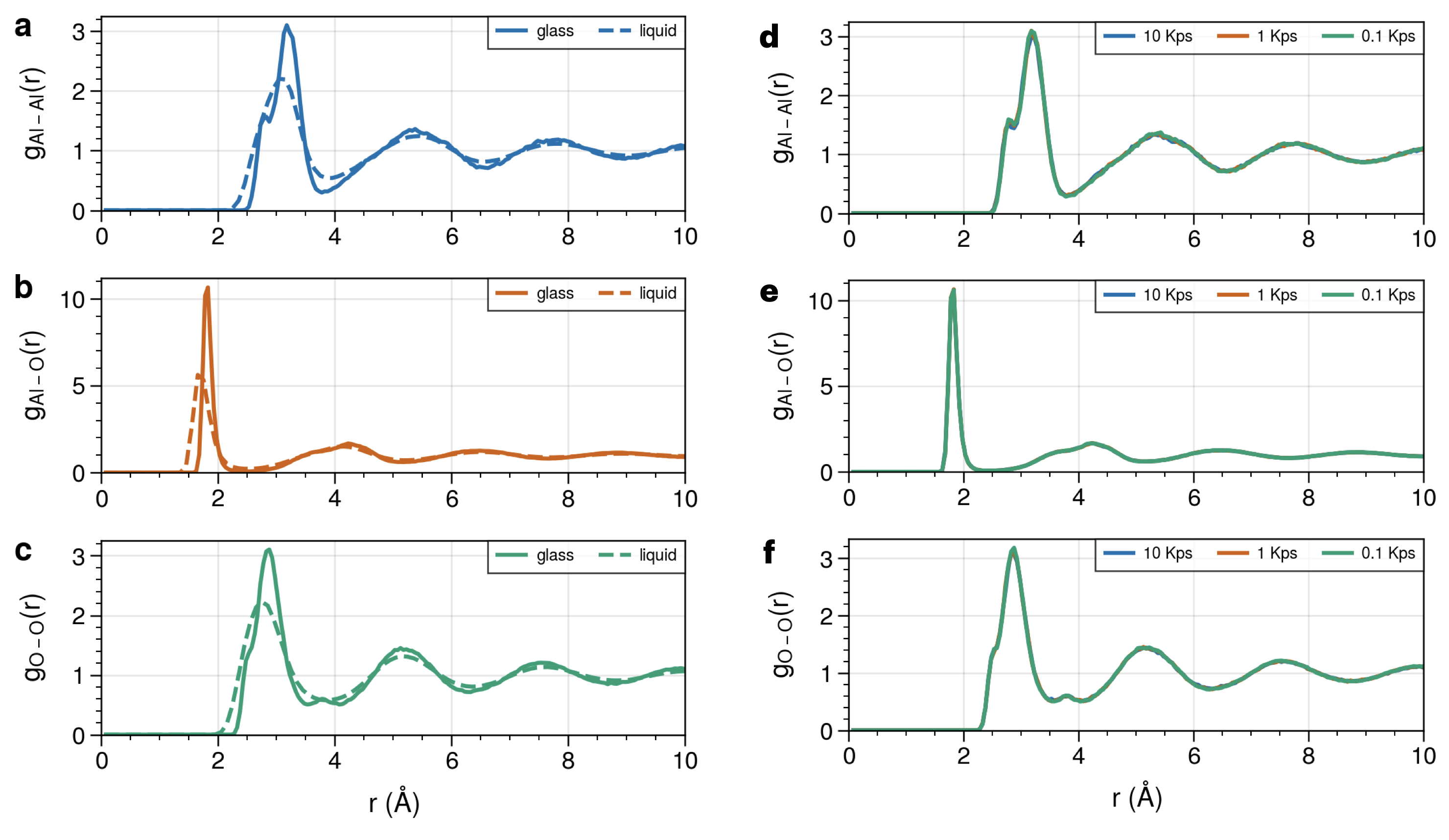}
    \caption{Partial radial distribution functions of bulk amorphous \alumina. (a--c) Glass (solid lines) compared with the equilibrium liquid at 2700~K (dashed lines) for Al--Al, Al--O, and O--O correlations; peak sharpening in the glass reflects reduced thermal broadening, while peak positions remain largely unchanged. (d--f) Comparison across cooling rates of 0.1, 1, and 10~K/ps; the near-complete overlap indicates that standard RDFs are insensitive to the structural differences induced by varying cooling rates.}
    \label{fig:prdf_glass}
\end{figure}

Figure~\ref{fig:prdf_glass}(a--c) compares the partial RDFs of bulk amorphous \alumina (glass) and the equilibrium liquid at 2700~K. The glass retains the same short-range order as the liquid, with sharpened peaks reflecting the loss of thermal broadening upon quenching. The Al--Al and O--O first peaks exhibit splitting in the glass, resolving distinct polyhedral connectivity motifs (e.g., edge- vs.\ corner-sharing) that are thermally broadened in the liquid. The Al--O first-peak position shifts slightly to larger distances, consistent with a change in the Al coordination number distribution upon quenching, as higher-coordinated species exhibit longer Al--O bond lengths.
To examine the effect of cooling rate on the short-range structure of the glass, Fig.~\ref{fig:prdf_glass}(d--f) compares the partial RDFs at three cooling rates (0.1, 1, and 10~K/ps). The RDFs are nearly indistinguishable, indicating that the standard pair correlation function alone is insufficient to characterize the subtle structural evolution over this range of cooling rates.

\section{Glass transition temperature}

\begin{figure}[hbtp]
    \centering
    \includegraphics[width=\linewidth]{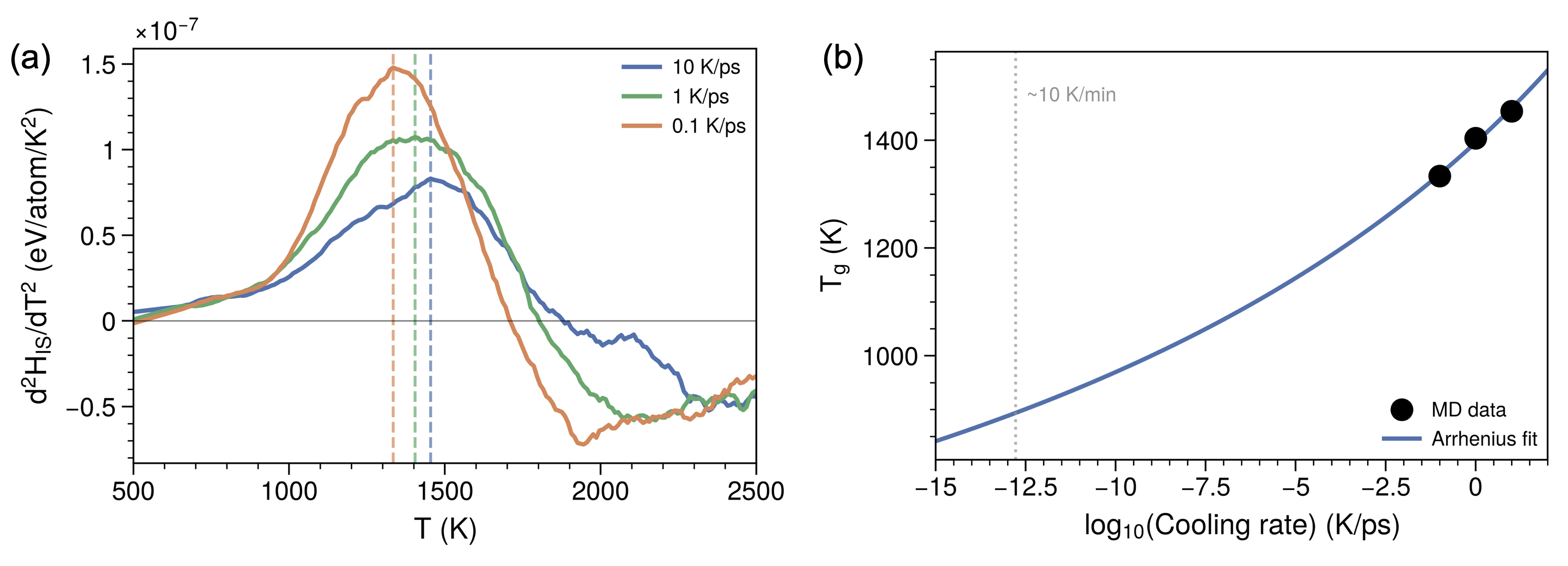}
    \caption{Glass transition analysis of amorphous \alumina from $H_\mathrm{IS}(T)$.
    (a) Second derivative $\mathrm{d}^2 H_\mathrm{IS}/\mathrm{d}T^2$ as a function of temperature for cooling rates of 0.1, 1, and 10~K/ps. Dashed vertical lines mark the peak positions used to determine $T_g$.
    (b) $T_g$ as a function of the logarithmic cooling rate $\log_{10}q$. Black circles denote the MD-derived $T_g$ values; the solid blue curve is an Arrhenius fit ($\ln q = A - E_a/R\,T_g$), corresponding to the $T_0 \to 0$ limit of the Vogel--Fulcher--Tammann relation. The vertical dotted line marks a representative experimental cooling rate of $\sim$10~K/min.}
    \label{fig:tg_vs_cr}
\end{figure}

The glass transition temperature $T_g$ was determined from the inflection point of the inherent structure enthalpy $H_\mathrm{IS}(T)$, identified as the temperature at which the second derivative $\mathrm{d}^2 H_\mathrm{IS}/\mathrm{d}T^2$ reaches its maximum absolute value during cooling.
Three cooling rates were examined: 0.1, 1, and 10~K/ps, each averaged over four independently generated configurations.
The resulting $\mathrm{d}^2 H_\mathrm{IS}/\mathrm{d}T^2$ curves are shown in Fig.~\ref{fig:tg_vs_cr}a, where the peak positions shift to higher temperatures with increasing cooling rate, yielding $T_g$ values of 1334~K at 0.1~K/ps, 1404~K at 1~K/ps, and 1454~K at 10~K/ps.
The cooling-rate dependence of $T_g$ is commonly described by the Vogel--Fulcher--Tammann (VFT) relation $\ln q = A - B/(T_g - T_0)$, where $q$ is the cooling rate.
However, fitting the present data yields $T_0 \approx 0$ over the accessible range of cooling rates, and the VFT relation reduces to the Arrhenius form $\ln q = A - E_a/(R\,T_g)$.
An Arrhenius fit is therefore applied for extrapolation (Fig.~\ref{fig:tg_vs_cr}b).
Extrapolation to a typical experimental cooling rate of $\sim$10~K/min gives $T_g \approx 894$~K, which lies within the experimentally measured range (743--943~K).\cite{heSolgelDerivedAlumina2019,hashimotoStructureAluminaGlass2022}

\section{Amorphous alumina surface}

\begin{figure}[hbtp]
    \centering
    \begin{tikzpicture}
       \node (a) {\includegraphics[width=0.7\textwidth]{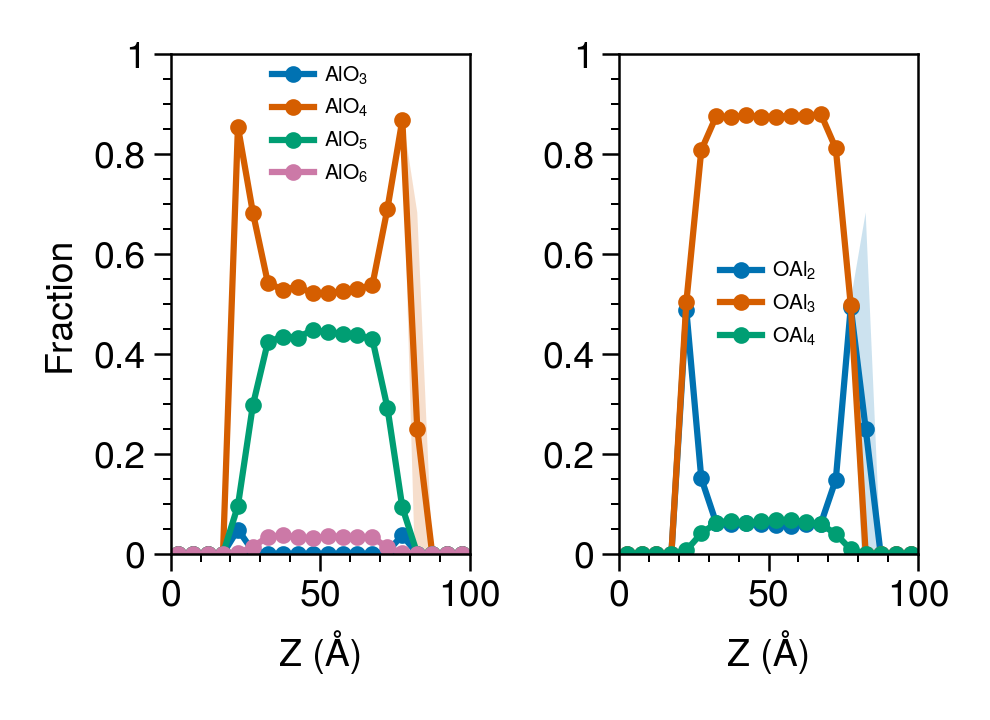}};
       \node at (-5cm,3.5cm) {\normalsize \textbf{a}};
       \node at (0.5cm,3.5cm) {\normalsize \textbf{b}};
    \end{tikzpicture}
    \caption{Coordination number profiles of amorphous \alumina films along the surface-normal ($z$) direction: (a) Al coordination fractions and (b) O coordination fractions. The two ends of the profile correspond to the free surfaces, while the central region represents the bulk-like interior. Results are averaged over four independently melt-quenched configurations generated using DP models trained with different random seeds; shaded regions indicate the standard deviation.}
    \label{fig:cn_profiles}
\end{figure}

Figure~\ref{fig:cn_profiles} shows the coordination number profiles of amorphous \alumina films resolved along the surface-normal ($z$) direction. In the bulk-like interior of the film, the Al coordination is dominated by AlO$_4$ ($\sim$54\%) and AlO$_5$ ($\sim$43\%) species, with minor contributions from AlO$_3$ and AlO$_6$, consistent with the bulk glass structure. Approaching the free surfaces, the AlO$_4$ fraction rises sharply to become the overwhelmingly dominant species at the outermost ridges, while AlO$_5$ is suppressed and AlO$_3$ rises from a negligible bulk-interior population to a clearly resolved peak at the surface, reflecting the truncation of the Al coordination shell at the vacuum interface.
The O sublattice exhibits a complementary but more dramatic redistribution: in the bulk interior OAl$_3$ ($\sim$88\%) overwhelmingly dominates over OAl$_2$ ($\sim$5\%), but at the outermost ridges the two species become nearly equally populated ($\sim$50\% each), reflecting the substantial loss of bridging oxygens at the free surface. OAl$_4$ remains a minor species throughout. This reorganization across only $\sim$10~\AA{} highlights how strongly the coordination environment of both sublattices is reshaped at the amorphous \alumina surface.

\begin{figure}[hbtp]
    \centering
    \includegraphics[width=0.5\linewidth]{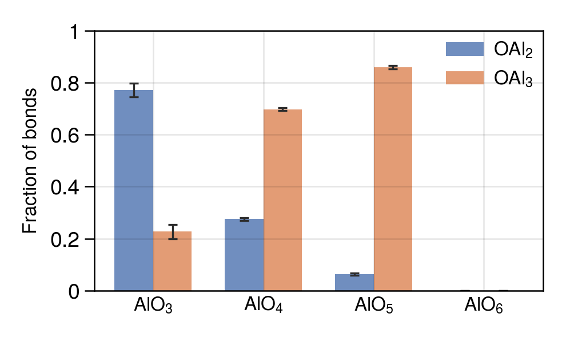}
    \caption{Comparison of the AlO$_n$ neighbor environments of OAl$_2$ (blue) and OAl$_3$ (orange) species at the amorphous \alumina free surface. For each surface oxygen, its bonded Al atoms are classified by coordination number (AlO$_3$, AlO$_4$, AlO$_5$, AlO$_6$); error bars denote $\pm 1$ standard deviation across four independent replicas. OAl$_2$ oxygens are strongly enriched in low-coordinated AlO$_3$ neighbors, whereas OAl$_3$ oxygens are predominantly bonded to higher-coordinated AlO$_4$ and AlO$_5$ sites, evidencing a spatial correlation between Al and O under-coordination at the surface.}
    \label{fig:OAl2neighbor}
\end{figure}

To quantify the spatial correlation between Al and O under-coordination at the amorphous \alumina free surface, we examined how the AlO$_n$ environments around OAl$_2$ and OAl$_3$ surface oxygens differ. For each surface OAl$_2$ (resp.\ OAl$_3$) site, we counted its bonded Al neighbors classified by their coordination number (AlO$_3$, AlO$_4$, AlO$_5$, AlO$_6$). As shown in Fig.~\ref{fig:OAl2neighbor}, the two oxygen species sample very different Al-neighbor environments. OAl$_2$ oxygens are strongly enriched in under-coordinated AlO$_3$ neighbors and depleted in higher-coordinated AlO$_5$ sites, while OAl$_3$ oxygens show the opposite preference, dominated by AlO$_4$ and AlO$_5$ neighbors. The almost mirror-image character of the two distributions demonstrates that low-CN Al and low-CN O preferentially share bonds at the free surface. This pairing reflects local bond-valence (Pauling) compensation: an under-coordinated OAl$_2$ oxygen recovers a near-bulk valence sum more readily by bonding to ``stronger'' low-CN AlO$_3$/AlO$_4$ partners (whose remaining Al--O bonds carry larger bond order), whereas higher-coordinated AlO$_4$/AlO$_5$ sites are best matched by more saturated OAl$_3$ oxygens.

\begin{figure}[hbtp]
    \centering
    \includegraphics[width=0.5\linewidth]{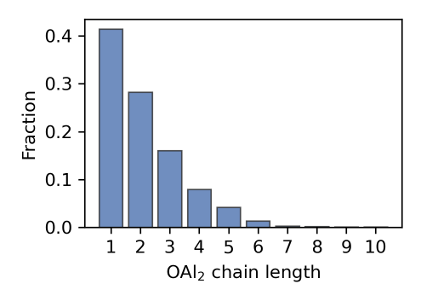}
    \caption{Distribution of connected-segment lengths for OAl$_2$ species at the amorphous \alumina free surface. Two OAl$_2$ oxygens are assigned to the same segment if they share a common Al neighbor; the chain length corresponds to the number of OAl$_2$ units in a connected segment. The distribution is dominated by isolated OAl$_2$ sites (length~1) and decays rapidly, with no segments longer than $\sim$6 units, indicating that surface OAl$_2$ motifs remain spatially fragmented rather than forming extended low-coordination sub-networks.}
    \label{fig:OAl2chain}
\end{figure}

The top-down view of the surface coordination map in Fig.~5a of the main text suggests, by visual inspection, that some OAl$_2$ sites may align into short chain-like arrangements. To assess whether such chains genuinely exist or are merely an apparent feature of the projection, we performed a quantitative connectivity analysis on the surface OAl$_2$ population. Two OAl$_2$ oxygens were considered to belong to the same connected segment if they share a common Al neighbor; the resulting clusters were then enumerated by their length (number of OAl$_2$ units). As shown in Fig.~\ref{fig:OAl2chain}, the distribution is strongly peaked at length one (isolated OAl$_2$ sites, $\sim$41\%), with monotonically decaying contributions from pairs ($\sim$28\%) and triplets ($\sim$16\%); segments longer than five units are essentially absent. Despite the large surface OAl$_2$ population, the under-coordinated oxygens therefore do not coalesce into extended chain-like or percolating sub-networks but remain spatially fragmented across the surface, indicating that the apparent linear features in Fig.~5a do not correspond to genuine OAl$_2$ chains.

\begin{figure}[hbtp]
    \centering
    \includegraphics[width=0.6\linewidth]{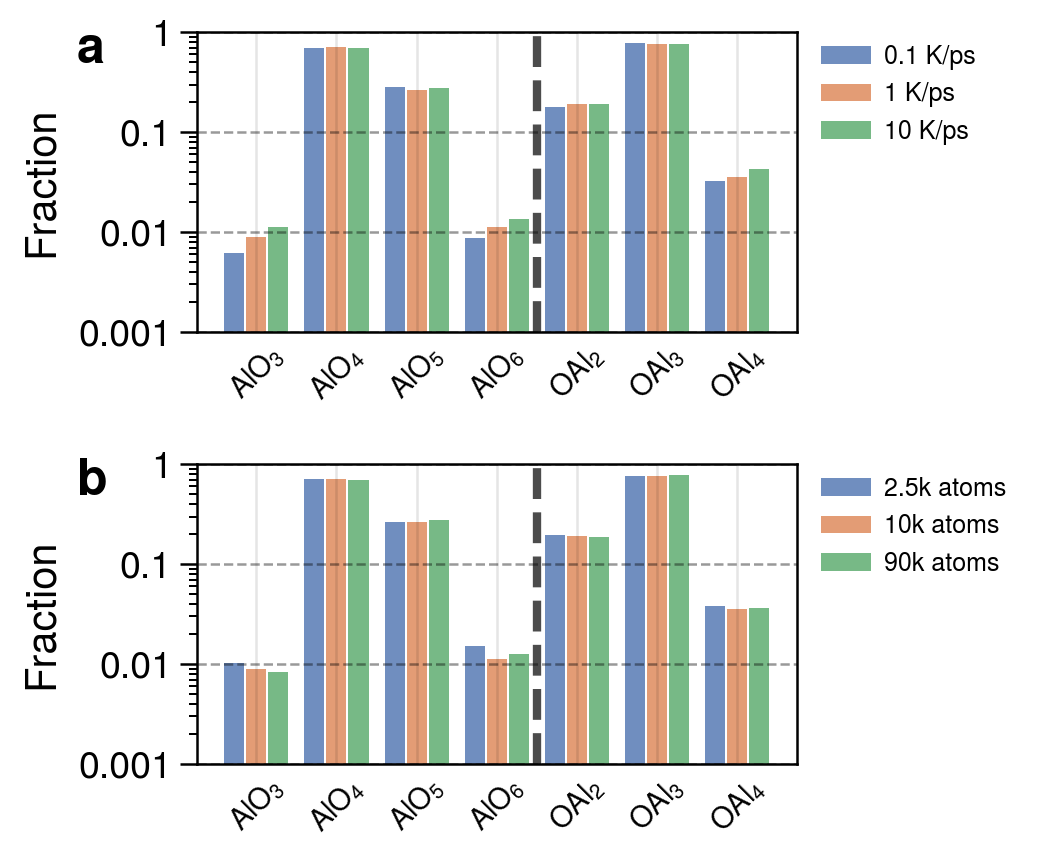}
    \caption{Dependence of surface coordination environments on cooling
    rate and system size using a geometric boundary. In contrast to Fig. 6 in the main text (which utilizes an intrinsic-interface construction), the surface atoms here are identified strictly by their Cartesian depth based on the $z$-resolved mass density profile shown in Fig. 3. 
    (a) Fractions of AlO$_n$ and OAl$_n$ species for a 10{,}000-atom system at three cooling rates (0.1, 1, and 10~K~ps$^{-1}$). Faster cooling consistently increases the populations of under-coordinated motifs (AlO$_3$, OAl$_2$). 
    (b) Corresponding fractions for three system sizes (2{,}500, 10{,}000, and 90{,}000 atoms) at a fixed cooling rate of 1~K~ps$^{-1}$. This depth-based definition confirms that the coordination fractions depend on the cooling rate (glass stability) but are less sensitive to system size.}
    \label{fig:CN_surf_effect}
\end{figure}

To ensure that our conclusions regarding system-size independence are not an artifact of the intrinsic-interface construction used in the main text, we performed an alternative coordination analysis using a static geometric boundary. In Fig.~\ref{fig:CN_surf_effect}, surface atoms are classified strictly based on their Cartesian depth, corresponding exactly to the ``surface'' region identified from the laterally averaged mass-density profile in Fig. 3b. As shown in Fig.~\ref{fig:CN_surf_effect}a, the sensitivity of the coordination distribution to thermal history remains clear: faster cooling generates a less relaxed surface with a higher fraction of under-coordinated Lewis acid (AlO$_3$) and Br\o nsted base (OAl$_2$) sites. Crucially, Fig.~\ref{fig:CN_surf_effect}b demonstrates that under this Cartesian definition, the coordination fractions are almost invariant across system sizes ranging from 2,500 to 90,000 atoms. This confirms that the statistical convergence of the surface chemistry is an intrinsic property of the simulated glass films and is entirely robust against the choice of surface definition.

\bibliography{alumina,xjy}